\documentclass[review,authoryear]{elsarticle}

\usepackage{amssymb,amsmath, bm,mathrsfs}
\usepackage{float}
\usepackage[ruled,vlined]{algorithm2e}
\usepackage{caption}
\usepackage{tabulary}
\usepackage{mdframed}
\usepackage{url}

\usepackage{lineno}
\modulolinenumbers[5]
\usepackage[hidelinks,colorlinks=true,citecolor=niceblue,linkcolor=niceblue]{hyperref}

\usepackage[margin=1in]{geometry}

\newmdtheoremenv{Definition}{Definition}
\newtheorem{proposition}{Proposition}
\newproof{pf}{Proof}
\newcommand{\R}{{}\operatorname{\mathbb{R}}}
\newcommand{\E}{{}\operatorname{\mathbb{E}}}

\newcommand{\Nat}{{}\operatorname{\mathbb{N}}}

\newcommand{\var}{\operatorname{Var}}

\newcommand{\defn}{\overset{\underset{\mathrm{def.}}{}}{=}}

\newcommand{\boldK}{\mathbf{K}}

\newcommand{\boldw}{\mathbf{w}}
\newcommand{\boldx}{\mathbf{x}}

\newcommand{\boldz}{\mathbf{z}}

\newcommand{\mcA}{\mathcal{A}}
\newcommand{\mcB}{\mathcal{B}}
\newcommand{\mcC}{\mathcal{C}}

\newcommand{\mcG}{\mathcal{G}}
\newcommand{\mcH}{\mathcal{H}}

\newcommand{\mcL}{\mathcal{L}}
\newcommand{\mcM}{\mathcal{M}}

\newcommand{\mcO}{\mathcal{O}}
\newcommand{\mcP}{\mathcal{P}}
\newcommand{\mcQ}{\mathcal{Q}}

\newcommand{\mcS}{\mathcal{S}}

\newcommand{\mcW}{\mathcal{W}}
\newcommand{\mcX}{\mathcal{X}}
\newcommand{\mcY}{\mathcal{Y}}

\newcommand{\argmin}{\operatornamewithlimits{argmin}}

\newcommand{\given}{ \mathop{\mid}}

\newcommand{\ident}{\mathbf{I}}

\newcommand{\trans}{^\top}

\newcommand{\iid}{\mathrel{\overset{\makebox[0pt]{\mbox{\normalfont\tiny\sffamily iid}}}{\sim}}}

\newcommand{\equivdist}{\mathrel{\overset{\makebox[0pt]{\mbox{\normalfont\tiny\sffamily dist}}}{=}}}

\newcommand{\wass}[3]{\mcW_{#3} \left( #1, #2 \right){}}

\newcommand{\avgwass}[3]{\overline{\mcW}_{#3} \left( #1, #2 \right){}}
\newcommand{\approxwass}[2]{\textsc{approxWass}\left( #1, #2 \right)}
\newcommand{\diag}{\operatorname{diag}}

\newcommand{\N}{\mathcal{N}}

\newcommand{\InvGaT}[2]{{}\operatorname{Inv-Gamma}\!\left(#1, #2\right){}}

\newcommand{\Un}{{}\operatorname{Unif}}

\newcommand{\Bern}[1]{\operatorname{Bern}\left( #1 \right)}

\begin{document}

	\title{Interpretable Model Summaries Using the Wasserstein Distance\tnoteref{t1}}
	\tnotetext[t1]{This document is the result of a research
		project funded by the National Institutes of Health.}
	\author[1,2] {Eric A. Dunipace\corref{cor1}
	}
	\ead{edunipace@mail.harvard.edu}
	\author[2,3] {Lorenzo Trippa}
	\address[1]{David Geffen School of Medicine at UCLA, 10833 Le Conte Ave, Los Angles, CA 90095, USA}
	\address[2]{Department of Biostatistics, Harvard T.H. Chan School of Public Health, 655 Huntington Avenue, Building 2, 4th Floor, Boston, MA 02115 USA}
	\address[3]{Department of Data Science, Dana-Farber Cancer Institute, CLSB 11045, 3 Blackfan Circle, Boston, MA 02115, USA}
	\cortext[cor1]{Corresponding author} 
	
	\maketitle
	
	\begin{abstract}
		Statistical models often include thousands of parameters.
		However, large models decrease  the investigator's ability to interpret and communicate the estimated parameters. Reducing the dimensionality of the parameter space in the estimation phase is a commonly used approach, but less work has focused on selecting subsets of the parameters for interpreting the  estimated model --- especially in settings such as Bayesian inference and model averaging. Importantly, many models do not have straightforward interpretations and create another layer of obfuscation. To solve this gap, we introduce a new method that uses the Wasserstein distance to identify a low-dimensional interpretable model projection. After the estimation of    complex models, users can budget how many parameters they wish to interpret and the proposed generates a simplified model of the desired dimension minimizing the distance to the full model. We provide simulation results to illustrate the  method and apply it to cancer datasets.
	\end{abstract}
	
	\section{Introduction}
	Statistical and machine learning models are changing many fields including medicine, politics, finance, and technology. Increasingly, these models diagnose diseases, predict elections, purchase stocks, and drive cars; however, interpreting  these models can be difficult. This difficulty is problematic because these models are used to make high stake decisions, sometimes to disastrous results
	\citep{Angwin2016}.
	Because of their increasing use, understanding how these models function, and diagnosing why they fail, is important. 
	
	Such models can be hard to understand because they function as black boxes, which  translate input data into predictions, but does ``not explain [its] predictions in a way that humans can understand" \citep{Rudin2018}.  
	Models fall into this black-box territory because of the method chosen---e.g. the coefficients in the third layer of an eight layer neural network do not have an intuitive interpretation---or because the model is so large as to be uninterpretable---like the logistic regression model discussed by \cite{Sur2019} that has 1,200 parameters.
	
	To remedy these failures, we propose summarizing black-box models with  low-dimensional and interpretable models that we call Sparse Local Interpretable Model (SLIM) summaries. 
	Our main contributions in this work are as follows.
	\begin{itemize}
		\item SLIM-a: a method to adaptively approximate the predictive distributions of \textit{any} model with a low dimensional interpretable linear model. This summary can be done in a local neighborhood around a single data point.
		\item SLIM-p: a method to summarize the predictive distributions of linear models in a way that \textit{preserves}  predictions from the original model. This interpretable summary eases interpretation and can be tailored  to clusters  of  samples or to a \emph{single data point.}
		\item Measures to monitor the accuracy of the SLIM approximations.
	\end{itemize}
	To do this, we utilize the $p$-Wasserstein distance to find low-dimensional interpretable models that produce predictions close to those of the original black-box model.  Additionally, our methods work in both frequentist and Bayesian settings to generate interpretable summaries relevant to the training data or for new predictor data entirely. We also provide code to replicate the paper and implement the method as an R package \citep{RCoreTeam2020} on GitHub at \url{https://www.github.com/ericdunipace/SLIMpaper}.
	
	These procedures could be useful in a variety of settings. For instance, a doctor may want to know the most important factors to measure in a patient to predict the patient's future risk of some disease or  an applied researcher may want to communicate parts of their model but not know which particular covariates to focus on. Alternatively, policy makers may want to understand how a complicated black-box model works.
	In all of these cases, SLIM is useful to \emph{slim} down the number of variables and increase interpretability.
	
	The rest of our paper is laid out as follows. First, we will define our notation and then describe the Wasserstein distance and its application in our setting in Section 2. Next, we describe a variety of computational methods that we have developed to estimate these interpretable summaries in Section 3. Section 4 provides simulation results demonstrating our method in a toy example,  a simple quadratic-linear model with Gaussian errors, and binary data with a  non-linear data generating process. Section 5 applies our method to a Bayesian Additive Regression Trees model estimating that estimates the survival time for glioblastoma patients, and, finally, Section 6 concludes.
	
	\section{Interpretable summaries using the Wasserstein distance}
	In this section, we will discuss several potential ways of summarizing statistical models. 
	The first is a SLIM-a method that estimates a new interpretable model to approximate the predictions from any model. Then we introduce a new method that we call SLIM-p that applies to linear models by finding the subset of covariates that generate predictions most faithful to the predictions from the original model. Before beginning, we define the notation and setting for the paper.
	
	\subsection{Setup and notation} \label{sec:setup}
	Suppose we have $n \in \Nat$ training observations, with predictor data $x_{1:n} = x_1,\ldots,x_n$ each in $\mcX \subseteq \R^k$ for $k \in \Nat$. We also have outcome data $y_{1:n} = y_1, \ldots, y_n$ each in $\mcY \subseteq \R$ and distributed according to density function $f_n$ with data-generating parameters $\theta \in \Theta$. Given $\{\{x_1,  y_1\}, \ldots, \{x_n,  y_n\}\}$, we estimate the parameters of this distribution via maximization or via sampling depending on our inferential framework. Let $\hat{\theta}$ denote these estimated parameters.
	
	In both frequentist and Bayesian settings, we may have a distribution of estimated parameters, though the underlying distributions rely on different philosophies. For frequentist estimation, the distribution of estimates arise via repeated sampling of data from the population of interest. Typically, this distribution is approximated using an asymptotic distribution or a bootstrap from $\{\{x_1,  y_1\}, \ldots, \{x_n,  y_n\}\}$. For Bayesian estimation, the distribution of estimates arise due to a prior distribution on the parameters that is updated by the likelihood. In either case, we assume that we have $T \in \Nat$ samples from the distribution of interest: $\hat{\theta}^{(1:T)} = \hat{\theta}^{(1)}, \ldots, \hat{\theta}^{(T)}$.
	
	Further, we have a function $m: \mcX \times \Theta \to  \mcH \subseteq \R$ that generates predictions where $ \mcH = g \left ( \E(Y_i \given X_i, \theta) \right)$ and $g: \R \mapsto \mcH$ is an invertible transformation like those in generalized linear models. 
	We will use this notation for inputs of both single vectors and sets of  vectors; however, the meaning should be clear from context. If we have a set of parameter vectors, then $m$ is a random function that maps to an empirical distribution on $\mcH$. If we have a set of data vectors of cardinality $N$, then $m$ projects to the space of $\mcH^N$ or $\mcH^{N \times T}$.
	
	For each observation $i$, we will have samples of the predictive distribution, \[m(x_i)  = \hat{\mu}_i^{(1:T)} = \begin{pmatrix} \hat{\mu}_i^{(1)} & \cdots & \hat{\mu}_i^{(T)} \end{pmatrix},\] each generated from $m$ for each value of $\hat{\theta}^{(1:T)}$ on $\mcH$. 
	We will refer to this quantity as the \textbf{\emph{original prediction}}. In a linear model,  $\hat{\mu}_i^{(1:T)} = \begin{pmatrix} x_i \trans \hat{\theta}^{(1)} & \cdots  & x_i \trans \hat{\theta}^{(T)} \end{pmatrix}$, but the original prediction does not have to be from a linear model. 
	To ease the notational burden, we will drop our subscripts when we wish to refer to the predictions for an entire set of data when the set of data is clear
	
	From this model, we will then generate a low-dimensional version from a class of interpretable models. Ideally, this will facilitate our ability to understand the original model. 
	Let $\hat{\nu}_i^{(1:T)} =  \begin{pmatrix} \hat{\nu}_i^{(1)} & \cdots & \hat{\nu}_i^{(T)} \end{pmatrix}$ be the samples from an interpretable model $q$ for observation $i$ with each $\hat{\nu}_{i}^t \in \mcH$. We will refer to this quantity as the \textbf{\emph{interpretable prediction}} because it relies on a reduced version of our predictor space. In this paper, all of the interpretable models we consider fall into the class of models that can be represented as a linear combinations of the $x_i^{(1:T)}$, i.e. $\hat{\nu}_i^t = \sum_{j=1}^k x_i^{(j)} \hat{\beta}_j$  with $\hat{\beta}_j^{(t)} \in \R$.

	\begin{table}[tb!]
		\begin{tabulary}{\linewidth}{cLc}
			\hline 
			Quantity & \centering{What is it?} & Where does it live?  \\ 
			\hline 
			$x_i$ & A $k$-length vector of predictors for observation $i$ & $\mcX \subseteq \R^k$ \\ 
			$y_i$ & Outcome data for observation $i$ & $\mcY \subseteq \R$\\
			$n$ & Number of observations in the training data & $\Nat$\\
			$\theta$ & Data-generating parameters for the outcome & $\Theta$\\
			$x_0$ & A $k$-length vector of predictors at which we want to interpret our model & $\mcX \subseteq \R^k$ \\ 
			$\mcB$ & A neighborhood around a point $x_0$  & $\mcB \subseteq \mcX$\\
			$N$ & Cardinality of $\mcB$ & $\Nat$\\
			$\mcP_T(\mcH)$ & The set of empirical distributions with $T$ elements over the space $\mcH$ & \\
			$m$ & A random function that generates predictions from the samples of parameter estimates from the primary model & $ \mcX\times \Theta^T  \mapsto \mcP_T(\mcH)$  \\
			$q$ & The random function that generates predictions from an interpretable model & $\mcX   \mapsto \mcP_T(\mcH)$  \\  
			$\hat{\mu}_i$ & The predictions from the random function $m$ generated with $x_i$. It is a distribution over $\mcH$. & $\mcP_T(\mcH)$  \\ 
			$\hat{\nu}_i$ & The predictions from $q(x_i)$. They form a distribution over $\mcH$ & $\mcP_T(\mcH)$  \\ 
			&  &    \\ 
			\hline 
		\end{tabulary}
		\caption{A table describing the important notation, what it is, and what space it is defined on.}
	\end{table}

	
	Finally, we have data $x_0$ for which we wish to interpret the model $m$. We define a neighborhood around $x_0$ as $\mcB = \{z \in \mcB  \subseteq \mcX: d(x_0, z) \leq \delta  \} $ for some $\delta > 0$ and distance metric $d$. We utilize this neighborhood $\mcB$ to identify our estimation of $q$, and to make our notation easier, we will abuse our notation by allowing functions defined in this paper to take in the neighborhood $\mcB$ as if it were a set of data. For instance, we define $m(\mcB) = \begin{pmatrix} m(z_1) & m(z_2) & \dots & m(z_N)  \end{pmatrix}\trans$ for all $z \in \mcB$. Let $N = |\mcB|$ be the cardinality of $\mcB$.

	With this notation, we now define a measure of closeness between the distributions of the full and interpretable predictions: the Wasserstein distance.
	
	\subsection{The Wasserstein Distance}
	The Wasserstein distance, also known as the Earth Mover's distance \citep{Rubner2000}, is a metric that can be used to measure the closeness between distributions. This distance metrizes the weak convergence between measures---i.e. convergence in distribution---and it can be used for measures that do not have the same support.
	For these reasons and others, the Wasserstein distance has found use in the statistical and machine learning literatures \citep{Peleg1989,Peyre2019, Bernton2019}.
	
	Let $d: \mcH^n \times \mcH^n \mapsto \R^+$ be a metric on the space $\mcH^n $, and let there be probability measures $\phi$ and $\eta$ defined on $\mcH^n $ each with finite $p^\text{th}$ moments.
	%
	%
	Then for empirical distributions $\hat{\phi}_{T} =\frac{1}{T} \sum_{t=1}^T \delta_{\mu^{(t)}}$ and $\hat{\eta}_{T'} = \frac{1}{T'}\sum_{t'=1}^{T'} \delta_{\nu^{(t')}}$ with all points in the space $\mcH^n$, the $p$-Wasserstein distance is defined as 
	\begin{equation}
		\mcW_p\left(\hat{\phi}_{T},\,\hat{\eta}_{T'} \right) = \inf_{\gamma \in \Gamma_{T,T'} } \left( \sum_{t=1}^T \sum_{t'=1}^{T'} d \left(\mu^{(t)}, \nu^{(t')} \right)^p \gamma_{t,t'} \right)^{\frac{1}{p}},
		\label{eq:wass_emp}
	\end{equation}
	where $\Gamma_{T,T'}$ is the set of all $T \times T'$ matrices with rows summing up to $1/T$ and columns summing up to $1/T'$. This matrix $\gamma$ specifies the joint distribution between samples denoted by $\mu$ and $\nu$ and is also known as the optimal transport plan. For this paper we will let $T = T'$ since both original and interpretable predictions will have the same number of atoms.
	%
	Additionally, we will write $\mcW_p\left(\hat{\phi}_{T},\hat{\eta}_{T} \right)$ in a slight abuse of notation as $\mcW_p\left( \hat{\mu}^{(1:T)}, \hat{\nu}^{(1:T)} \right)$, similar to \cite{Bernton2019}. 
	
	For the predictions $\hat{\mu}^{(1:T)}$ and $\hat{\nu}^{(1:T)}$, the Wassersten distance serves as a measure of discrepancy between the two distributions and how much predictive accuracy is lost by using $\hat{\nu}^{(1:T)}$ instead of $\hat{\mu}^{(1:T)}$. 
	With $T=1$ and with the Euclicean metric, the 2-Wasserstein distance is equal up to a constant to the root-mean-squared error. Indeed, \cite{Villani2006} recommends thinking of the 2-Wasserstein distance as the minimal Euclidean distance between sets of points.
	%
	
	We pause for moment to consider some of the advantages of the Wasserstein distance over other distance metrics. AS mentioned previously, the Wasserstein distance metrizes the convergence in distribution and, unlike Total Variation distance or the Chi-Square distance, the Wasserstein distance is defined for disjoint measures. This is usually the case with empirical samples from separate distributions, and will be the case in the setting of this paper especially when considering Bayesian methodology. Third, the Wasserstein distance is a true distance, \textit{i.e.} the triangle inequality holds:  $\mcW_p\left(A,\,B \right) \leq \mcW_p\left(A,\,C \right) + \mcW_p\left(C,\,B \right)$. Fourth and finally, there is nothing special about the Wasserstein distance for interpretability \textit{per se}; instead, it is a tool that will allow us to measure how close two predictive distributions actually are, one distribution being from the original prediction and the second being from an easier to interpret model.
	With this in mind, we turn to approximating predictions with sparse local interpretable model summaries.

	\subsection{Sparse local interpretable model-agnostic summaries}\label{sec:slima}
	We have developed two flavors of interpretable approximations: model agnostic and model preserving. To distinguish these two flavors of SLIM we call the model-agnostic version SLIM-a and the model-preserving version SLIM-p. First, we will discuss the SLIM-a version of our summaries. In short, the basic idea is to fit a model that is easy to understand, such as a linear regression, to the predictions from a more complicated model. This method will help explain how the input data translate to predictions, at least approximately. 
	
	We consider three general aspects when finding an interpretable approximation before discussing SLIM-a directly. Those aspects are the closeness between the original and interpretable models, the dimensionality of the interpretable model, and global versus local approximations.  First, predictions from the interpretable approximation should be close to the predictions from original model, which means the interpretable model $q$ generates predictions similar to those of $m$. Second, the interpretable model should also be low dimensional to facilitate understanding. Indeed, even interpretable models effectively become black boxes when the dimension is large since humans can only focus on about seven pieces of information at once \citep{Miller1955}. These criteria become important when distinguishing between global and local interpretable models.
	
\begin{figure}[!tb]
		\includegraphics[width=\textwidth]{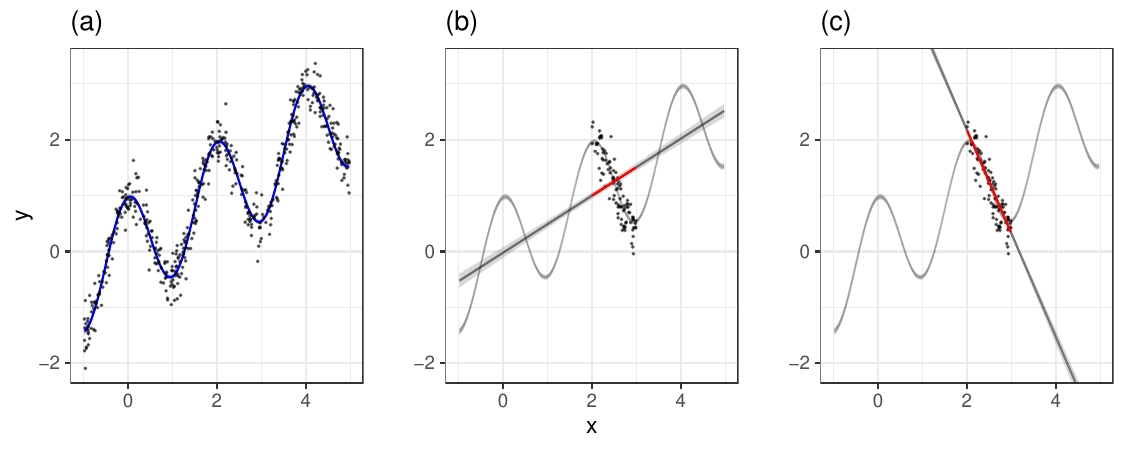}
		\caption[Global versus local interpretable approximations for a ``black-box model'']%
		{Global versus local interpretable approximations for a ``black-box model.'' Data generated from $X \sim \Un(-1,5)$ and $Y \sim \N(\cos(\pi X) + 0.5 X, 0.25^2)$ with 512 observations. The left most plot, (a), shows the original data and the predicted mean from a posterior fit using the true data generating model. Independent standard Normal priors were used for the coefficients. The middle plot, (b), shows a sample from a subset of the predictor space, $X \in (2,3)$ and a global interpretable approximation (an ordinary least squares regression) made by regressing predictions from the non-linear model, $\hat{Y}$, on all of the predictor data. The portion relevant to the subsample is highlighted in red. The final plot, (c), shows the result from using a local interpretable approximation based on regressing the predictions from the non-linear model on the subset of the predictor data with the relevant portion again highlighted in red. In both (b) and (c) the lines extend past the region of interest to more clearly show the slopes of the lines and 95\% credible intervals.}
		\label{fig:nonlinear_approx}
	\end{figure}

	Third, the scales of explanation vary between global versus local interpretable models. Global interpretable models are explanatory models fit to the predictions from an entire data set while local interpretable models focus on a subset of the predictor space. Variables that are useful in global interpretable models would likely also be important to generate local predictions; however, a global interpretable model will not generally be close to the original model.
	Alternatively, local interpretable models are more likely to be close to the predictions for the relevant subset of data. While this model may give a good description of how inputs translate to predictions locally, i.e. be close to the original model $m$,  the interpretable model may perform quite poorly outside of this subset and give predictions that are quite far from the original model.  Ideally, the interpretable approximation would achieve predictions close to those of the black-box model, but with the added benefit that it is possible to understand how the interpretable model behaves.  Figure \ref{fig:nonlinear_approx} provides a toy example of a global and local interpretable model in a Bayesian setting. 

	To achieve an interpretable approximations in the SLIM-a framework, we propose minimizing the loss between the original model and predictions from an interpretable model---denoted as $q$---from a class of interpretable models $\mcQ$, given a neighborhood around the  predictors for a particular observation
	$\mcB$. We will define this neighborhood more precisely shortly.
	Further, we also introduce a penalty function $P$ on the complexity of the interpretable model $q$. Then we seek to find
	\begin{equation*}
		l(x_0) = \argmin_{q \in \mcQ} \mcL(m, q, \mcB) + P(q),
		\label{eq:interp_ob}
	\end{equation*}
	utilizing a modified version of the notation of \citet{Ribeiro2016} We utilize a loss function of
	\[\mcL(m, q, \mcB) = \wass{m\left[\mcB\right]}{q \left[\mcB\right]}{p}^p.\] 
	
	For purposes of explication, we assume that $T=1$ and that we are focused on a single test observation  $x_0$ such that $m: \mcX \mapsto \mcH^{1 \times 1}$. 
	Taking $\mcQ$ to be the class of generalized linear models, the linear prediction function $q$ on $\mcH$ is 
	\[
	q(x_0) = x_0 \trans \beta .
	\]
	As a reminder, we are working on the \textit{linear predictor space} of the interpretable model, and not the conditional mean space. Hence, $q$ will be a linear function in this space.
	Since $T=1$, the loss will be
	\[ 
	\mcL(m, q, \mcB) = \sum_{z \in \mcB}\left \| m(z) - z\trans \beta \right \|_p^p.
	\]
	
	
	With samples of values of $\hat{\theta}$, our $m$ will generate $T \geq 1$ predictions for each observation.
	Further, to make the distribution generated by $m$ more explicit, we will write $m(z) = \hat{\mu}_z^{(1:T)}$ for $z \in \mcX$. 
	Thus, we will utilize the loss function
	\begin{equation}
		\mcL( \hat{\mu}^{(1:T)}, \beta^{(1:T)}, \mcB) =  \inf_{\gamma \in \Gamma_{T,T'} } \sum_{z \in \mcB} \sum_{t=1, t' =1}^{T} \| \hat{\mu}_z^{(t)} - z \trans \beta^{(t')}\|_p^p \gamma_{t,t'}
		\label{eq:full_loss}
	\end{equation}
	with each $\beta^{(t')} \in \R^k$. We write $\beta^{(1:T)}$ to explicitly refer to all $k \times T$ values of the coefficients. Further, this loss function readily admits a penalty like $P(q) = \lambda \sum_{j=1}^k \|\beta_j^{(1:T)}\|_2$, which is a group LASSO on the dimensions of $\beta^{(1:T)}$ \citep{Yuan2006}.
	
	With the loss and penalty functions defined, we now discuss the selection of the neighborhood for $x_0$.
	
	The neighborhood around $x_0$ can be defined in several ways. One is that the neighborhood could be user-defined interpretable changes from the current value of $x_0$ such that the values of $z$ have meaning for the applied researcher. For example, one could compare the observed $x_0$ to a pseudo-observation with similar characteristics but with a one-unit change on a variable such as age. Another potential way to define a neighborhood is to take some number of observations in the data closest to $x_0$, which gives a sense of how the prediction function changes for the observed values close to $x_0$. Finally, the neighborhood could be points simulated from a probability distribution, such as Gaussian a distribution with mean $x_0$ and covariance equal to a scaled version of the empirical covariance of the data used to estimate $m$, such as $\hat{\Sigma}/n$. 
	
	Note well that these neighborhoods also define the scale of the interpretable summaries. If we take our neighborhood to be $x_{1:n}$, then the interpretation becomes how the prediction function changes on average across all the data, i.e. the global model in Figure \ref{fig:nonlinear_approx}. Similarly, we can make our neighborhood much closer to the values of $x_0$, which is the local model in Figure \ref{fig:nonlinear_approx}. As we decrease the neighborhood to its limit, then the local approximation becomes an approximation of the prediction function derivative at $x_0$ (see Section \ref{sec:bin_eg} for an example). This can be achieved as the Gaussian-based neighborhood achieves a degenerate variance of 0. There may be reasons to prefer one of these neighborhoods over another, but we will explore the last of these---the Gaussian neighborhood---in our simulation studies in Section \ref{sec:sim}. An additional point is that for the identification of the parameters $\beta^{(1:T)}$ in the SLIM-a model, $N$, the cardinality of $\mcB$, should be greater the number of covariates in the interpretable model.
	
	We also note that the term inside the infimum looks like a multivariate $L_p$ regression with different ``outcomes'' for each observation being the $T$ samples of the original prediction. This then means that for a given $t' \in \{1,\ldots,T\}$,  $\beta^{(t')}$ will minimize the $L_p$ distance to the corresponding $\mu^{(t)}$ without any need for the mass matrix $\gamma$. 
To see this, note that the atoms of each distribution have the same number of points, $T$, which means $\gamma$ will be a permutation matrix \citep{birkhoff_three_1946}. Then since $\beta^{(t)}$ minimizes the $L_p$ distance to $\hat{\mu}_z^{(t)}$,  $\| \hat{\mu}_z^{(t)} - z \trans \beta^{(t)}\|_p^p \leq  \| \hat{\mu}_z^{(t)} - z \trans \beta^{(t')}\|_p^p \, \forall t' \neq t.$ Thus, the permutation matrix that pairs $t = t'$ is a minimizer. 
	Hence we drop the infimum from Eq. (\ref{eq:full_loss}) and use the loss function
	\begin{equation}
		\mcL( \hat{\mu}^{(1:T)},\beta^{(1:T)}, \mcB) = \sum_{z \in \mcB} \sum_{t=1}^T \| \mu_z^{(t)} - z \trans \beta^{(t')}\|_p^p.
		\label{eq:loss_l2}
	\end{equation}	
	Finally, combining this modified loss with the group LASSO penalty, the objective function is
	\begin{equation}
		l(x_0) = \argmin_{\beta^{(1:T)}} \mcL( \hat{\mu}^{(1:T)},\beta^{(1:T)}, \mcB) + \lambda \sum_{j=1}^k \left \|\beta_j^{(1:T)} \right \|_2.
		\label{eq:obj_l2}
	\end{equation}
	
	This concludes our description of the model-agnostic approach for SLIM. In the case where $m$ is also a generalized linear model, one may instead desire an approach that is model-preserving---that is where the values estimated by the interpretable model are the same as the coefficients estimated in the original model. A researcher may desire this when he or she is simply seeking to decide which coefficients to report from a large regression model without changing the coefficients themselves. Our tool for such a setting  is the one we describe next.
	
	\subsection{Sparse local interpretable model-preserving summaries for linear models} \label{sec:slimp}
	
	The previous section was model-agnostic since the original functional form of the model generating the original predictions had no bearing on the class of interpretable models considered. We propose a modification of this model-agnostic concept, which we call SLIM-a, to include model-preserving local interpretable explanations for linear models or SLIM-p.
	
	Generalized linear models are perhaps the most widely used statistical models owing to their ease of implementation and interpretability. For these models, $m(x_0) = \sum_{j=1}^k x_0 \hat{\theta}_j$ and we write $m(x_0) = \hat{\mu}_0^{(1:T)}$ to emphasize the fact that these predictions are a distribution of $T$ empirical samples whether in a frequentist or Bayesian context. After variable selection and the estimation of a particular model, practitioners may in some sense believe their model is true, or at least trust that the values of the coefficients coming from a model are believable. In these cases, the SLIM-a method may not be desirable since it will modify the existing coefficients. Instead, we can restrict the class of interpretable models to $\mcQ ^ \ast  = \{ q \in \mcQ: \beta_j = \theta_j\}$. In words, $\mcQ ^\ast$ is the set of models that use the same values for the coefficients as found in $m$. We can then let $\beta_j$ for the interpretable model be equal to $\alpha_j \hat{\theta}_j^{(1:T)} $, where $\alpha_j$ is a binary indicator if component $\hat{\theta}_j$ is on or off. Then the interpretable model $q$ would be 
	\begin{equation}
		q(x_0) =\sum_{j=1}^k x_0^{(j)} \alpha_j \hat{\theta}_j^{(1:T)}  , \text{ s.t. } \alpha \in \{0,1\}^k.
		\label{eq:SLIM-p}
	\end{equation}
	The penalty term in this case would be $P(\alpha) = \lambda \sum_{j=1}^k \alpha_j$, or the number of non-zero elements of $\alpha$ times $\lambda$.
	
	Substituting the values of $\beta_j = \alpha_j \hat{\theta}_j$ into Eq. (\ref{eq:full_loss}), we get a loss of
	\begin{equation*}
		\mcL( \hat{\mu}^{(1:T)},\alpha, \mcB) = \inf_{\gamma \in \Gamma_{T,T'} } \sum_{z \in \mcB}  \sum_{t=1, t' =1}^{T} \left \| \mu_z^{(t)} - z \trans \diag(\alpha) \hat{\theta}^{(t')} \right \|_p^p \gamma_{t,t'},
	\end{equation*}
	where $\diag(\cdot)$ turns a vector into its corresponding diagonal matrix. In this case, $\alpha$ is identified when the neighborhood $\mcB = \{x_0\}$. So the loss becomes
	\begin{equation}
		\mcL( \mu,\alpha, x_0) =\inf_{\gamma \in \Gamma_{T,T'} } \sum_{t=1, t' =1}^{T} \left \| \mu_0^{(t)} - x_i \trans \diag(\alpha) \hat{\theta}^{(t')} \right \|_p^p \gamma_{t,t'}.
		\label{eq:slimp_loss}
	\end{equation}
	When we combine this loss with the penalty $P(q)=\lambda \sum_{j=1}^k \alpha_j$, we obtain an objective function of 
	\begin{equation*}
		l(x_0) = \argmin_\alpha \inf_{\gamma \in \Gamma_{T,T'} } \sum_{t=1, t' =1}^{T} \left \| \mu_0^{(t)} - x_i \trans \diag(\alpha) \hat{\theta}^{(t')} \right \|_p^p \gamma_{t,t'}+ \lambda \sum_{j=1}^k \alpha_j.
	\end{equation*}
	
	Note that we can also define an interpretable model for a collection of individuals for a neighborhood $\mcB$ around $x_0$:
	\begin{equation}
		l(x_{0}) = \argmin_\alpha \inf_{\gamma \in \Gamma_{T,T'} }  \sum_{z \in \mcB} \sum_{t=1, t' =1}^{T}\left  \| \mu_z^{(t)} - z \trans \diag(\alpha) \theta^{(t')} \right \|_p^p \gamma_{t,t'}+ \lambda \sum_{j=1}^k \alpha_j.
		\label{eq:slimp_obj}
	\end{equation}
	Again, the neighborhood can be defined for whatever interpretation is desired. Also, remember that the neighborhood for SLIM-p models can consist of a \emph{single} observation. This is identified due to the constraints of the model.
	We defer the computational aspects of calculating these quantities to Section \ref{sec:comp}.
	
	\subsection{Monitoring performance}
	The use of the $p$-Wasserstein distance facilitates an easy check of performance for our interpretable models since it, by definition, measures how close two distributions are. Conceptually, the $p$-Wasserstein performs like the distance between vectors of predictions from the interpretable model and the original model. However, we may also be curious about how bad or how good the prediction may be on a set of data relative to the full model. Towards this end, we propose 3 ways to measure performance.
	
	\paragraph{1. The $p$-Wasserstein distance} The obvious metric to measure performance is the the $p$-Wasserstein distance between predictions from $m$ and $q$. However, this metric is not necessarily intuitive so some facts about the distance are helpful to make sense of it. For example, when the $p$-Wasserstein distance is 0,  the two models generate the same predictions. Another fact for the SLIM-a case is that the largest distance will be from a null model $q_{\text{null}}$, usually for an intercept only model or when all coefficients, including the intercept, are set to 0. From this fact, we can create another measure of performance.
	
	\paragraph{2. The Wasserstein $R^2$} The $R^2$ is constructed as 
	\[1- \frac{\E(\var(Y \given X))}{\var(Y)},\] which gives a measure of the percent of variance explained by a model since $0 \leq \E(\var(Y \given X)) \leq \var(Y)$. We have a similar relationship $p$-Wasserstein distances using an appropriately chosen $q_{\text{null}}$  and any other $q$: $0 \leq \wass{m}{q}{p} \leq \wass{m}{q_{\text{null}}}{p}.$ 
	From this, we define the Wasserstein $R^2$.\\
	
	\begin{Definition}[The $p$-Wasserstein $R^2$]\label{def:wassr2}
		For models $m$, $q$, and $q_{\text{null}}$ and a neighborhood $\mcB $ around an observation $x_0$, the $p$-Wasserstein $R^2$ is
		\[ 
		1 - \frac{\wass{m\left[\mcB \right]}{q \left[\mcB \right]}{p}^p}{\wass{m\left[\mcB\right]}{q_{\text{null}} \left[\mcB \right]}{p}^p},
		\]
		with values of $\frac{\wass{m\left[\mcB \right]}{q \left[\mcB \right]}{p}^p}{\wass{m\left[\mcB\right]}{q_{\text{null}} \left[\mcB \right]}{p}^p} = \frac{0}{0}$ defined as 0 and values of $\frac{\wass{m\left[\mcB \right]}{q \left[\mcB \right]}{p}^p}{\wass{m\left[\mcB\right]}{q_{\text{null}} \left[\mcB \right]}{p}^p} = \frac{d}{0} \,\, \forall d \in \R^+$ defined as $\infty$.
	\end{Definition}
	The Wasserstein $R^2$ is bounded between $[0,1]$ like for the $R^2$ in traditional linear models if the null model correctly chosen. For example, $q_null$ in our interpretable model setting could be a model with just an intercept.
	Moreover, this quantity gives a better intuitive sense of how close the two distributions are. A value of 0.9 indicates that the predictive distribution of the current interpretable model $q$ has closed 90\% of the $p$-Wasserstein distance to the original model as compared to the predictive distribution of the minimal model $q_{\text{min}}$.
	
	\paragraph{3. Average Wasserstein distances} We may also desire a measure of how close the predictions are on average for each observation in the set of data we wish to interpret, $x_0$. Towards this end, we propose an average $p$-Wasserstein distance.
	
	\begin{Definition}[The average $p$-Wasserstein distance]\label{def:avgwass}
		For a set of data we wish to interpret, $\mcB$, the average $p$-Wasserstein distance is
		\[\avgwass{m\left[\mcB \right]}{q \left[\mcB \right]}{p} = \frac{1}{N} \sum_{z \in \mcB}\wass{m(z)}{q(z)}{p}\]
	\end{Definition}
	
	From this, we can also select individuals $i$ or $i'$ with the best and worst performing individual p-Wasserstein distances to see both how well, and how poorly the predictions may perform. We can also select any quantile such as the median or the $25^{\text{th}}$ percentile. In the case where we generate predictions for a single individual, these quantiles will obviously all be the same.
	
	\subsection{Previous work}
	Our work bares similarity to previous work in the literature. Conceptually, the SLIM-a method is similar to the Kullback-Leibler projection method developed by \cite{Goutis1998}  and \cite{Dupuis1998}, further developed by \cite{Nott2010}, and as described by \cite{Piironen2015} and \cite{Piironen2017}. Unlike these authors, we do not solely use Bayesian methods nor is our goal model selection.  However, Eq. (\ref{eq:obj_l2}) will be equivalent to their projections when the outcome in Gaussian and the methods are Bayesian.
	
	Other work in the field of interpretable summaries of Bayesian linear models includes \cite{Hahn2015}. The authors of this paper utilize the posterior expectation of predictions to obtain a sparse version of the posterior mean over the coefficients. However, this method has the disadvantage that it only returns the posterior mean rather than a full distribution over the non-zero coefficients. In contrast, the proposed method in this paper is distribution preserving.
	
	Another paper by \cite{Ribeiro2016} discuss using interpretable models to explain predictions from any arbitrary model in frequentist settings, a method they call local interpretable model-agnostic explanation or LIME. They minimize a Euclidean loss function between predictions from a model and an interpretable model such as a linear regression. Their method is conceptually similar to the SLIM-a method with a single posterior sample. \cite{Woody2019} applied the method of \citeauthor{Ribeiro2016} in a Bayesian setting also using a Euclidean loss. However, none of these papers discuss the SLIM-p method nor do they develop the theoretical justification for these interpretable projections.
	
	Of course, the field of model summarization is an active field. \cite{Doshi-velez2018} discuss the criteria by which interpretability should be defined and how to evaluate these interpretations---for example through predictive accuracy and ease of human understanding---though they do not discuss summaries directly. Other authors like \cite{strumbelj_explaining_2014} focus on perturbing the input variables to see how much predictions change. In a similar vein, various authors have discussed measuring variable importance as a means to interpret model, \textit{i.e.} more important variables are perhaps more important to predictions. However, there is no unique mathematical definition of variable importance and authors have used the word to mean a variety of things. See \cite{Gelman2008}, \cite{Koh2017}, and \cite{Shrikumar2017} for some examples of the uses in the literature. Even authors like \cite{Datta2016} can be seen as providing a method that measures variable importance. 
	
	Other work focuses on methods that highlight portions of an input image. \cite{bach_pixel-wise_2015} identify the contributions of single pixels to the predictions of a machine learning model by generating heatmaps of the input images that relate to the pixels contributions to the model.
	\cite{Zintgraf2017} similarly discuss highlighting the various parts of an image that ``provides evidence for or against`` an image belonging to a certain class.
	
	In contrast to these attempts at model explanation, work by \cite{Rudin2014, Rudin2018} discuss the dangers of simply trying to explain predictions from models with some hilarious examples of explanations gone wrong. This work also helpfully defines what is meant by black-box models.
	In the same vein, \cite{Chen2018} discuss an applied example making credit scores more interpretable.
	
	Other papers, like this one, instead focus using simplified models to interpret more complicated ones. \cite{lundberg_unified_2017} provide a framework to interpret models using a measure of variable importance and additive models on the covariates.
	And a recent paper by \cite{agarwal_neural_2021} proposes a similar idea to this paper but using Neural Additive models which trains sub-neural networks on each covariate.

	Unlike the above work, SLIM offers several advantages. We provide a connection between the limiting Frequentist case where the distribution of parameters is degenerate (a single estimate) to the case where we have a distribution of possible estimates from the hypothesized sampling distribution and further to the Bayesian case where the samples are from a distribution on the parameters. Additionally, SLIM allows for powers other than the Euclidean distance and hence it is more general. Moreover, the complexity of the method scales with that of LASSO, which can be applied to large data sets \citep{suchard_massive_2013}. Finally, the Wasserstein distance gives useful diagnostics for how well the approximate model performs in terms of the distribution of estimates. 
	
	
	With these three interpretable methods---variable importance, SLIM-a, and SLIM-p---and discussion of previous work, we now turn to the computational methods used to generate these quantities.
	
	\section{Computing interpretable summaries}\label{sec:comp}
	In the previous section, we discussed several measures to interpret statistical models: variable importance, SLIM-a, and SLIM-p. We now discuss several algorithms to estimate our interpretable measures. But first, we discuss how to estimate the transportation matrix, $\gamma$.
	
	\subsection{Optimal transport plans and approximate Wasserstein distances} \label{sec:ot_est}
	Many methods exist to calculate the transportation matrix utilized in Eq (\ref{eq:wass_emp}). For univariate distributions, one simply has to sort both empirical distributions and match points across distributions by their ranks (see remarks 2.28 and 2.30 in \citealp*{Peyre2019}). Outside of univariate distributions, algorithms exist to calculate optimal transport plans in $\mcO(T^3)$ \citep{Solomon2015}, such as the shortlist method of \cite{Gottschlich2014} implemented in the \texttt{transport} R package \citep{Schuhmacher2019}.
	
	Due to the poor computational performance of exact optimal transport, regularized optimal transport has found increasing use in the literature in recent years.
	\cite{Cuturi2013} uses a negative entropy penalty on the transport plan $\gamma$ while \cite{Bernton2019} utilize Hilbert space-filling curves to find approximate optimal transport solutions \citep{Hilbert1891}. In this work we utilize the exact algorithm of \citeauthor{Gottschlich2014}, but did not find appreciable difference in performance by using approximation algorithms in our experiments.
	
	To estimate $\gamma$, 
	we define a general function $\text{OT}$ that takes in samples of $\mu$ and $\nu$ and then returns a transport matrix $\gamma$:
	\begin{equation}
		\hat{\gamma} = \operatorname{OT}(\hat{\mu}, \hat{\nu}).
		\label{eq:OTfun}
	\end{equation}
	Though we only use exact optimal transport plans, we still define an algorithm that can generate a plan for an exact or approximate $p$-Wasserstein distance.
	\begin{algorithm}[!hbt]
		\KwData{original prediction  $\hat{\mu}$ in $\mcH^{N \times T}$ and interpretable prediction $\hat{\nu}$ in $\mcH^{N \times T}$
		}
		\KwResult{distance $\hat{d}  \in \R^+$}
		Set $\hat{\gamma} =  \operatorname{OT}({\mu}, {\nu})$\;
		Set $\hat{d} = \sum_{t=1, t'=1}^T \left \| {\mu}^{(t)} - {\nu}^{(t')} \right\|_p^p \hat{\gamma}_{t,t'}$\;
		\KwResult{$\hat{d} $}
		\caption{\textsc{approxWass}($\mu$, $\nu$)}
		\label{alg:approxWass}
	\end{algorithm}

	With these details about calculating optimal transport plans, we can now discuss the estimation of variable importance, SLIM-a, and SLIM-p.
	
	\subsection{SLIM-a estimation}
	The ease of estimating Eq. (\ref{eq:obj_l2}) depends on the chosen power of the $p$-Wasserstein distance. 
	
	For $p = 2$, the problem is equivalent to a multivariate least-squares regression with an $L_1$ penalty. One popular choice to estimate such models is the  \texttt{glmnet} package in R that allows for multivariate linear regression with an $L_1$ or elastic net penalties \citep{Friedman2010}. However we instead utilize a modified version of the \texttt{oem} package in R \citep{Huling2018}. The package  allows for sparse regression with a variety of penalties using the orthogonalizing expectation maximization algorithm of \cite{Xiong2016}. This algorithm has the benefit that it is relatively fast and that it will converge to a regression using the pseudo-inverse when the design matrix is not full rank; however, the original software could only handle single outcome regression, so we have extended the program to incorporate multivariate linear regression.
	
	We also consider the special cases where $p \in \{1, \infty \}$. For $p=1$, we rely on a modified version of the \texttt{rqPen} package that performs group LASSO in the case of regression with an $L_1$ loss \citep{Sherwood2020}. The case of $p = \infty$ is equivalent to minimizing the maximum residual and we rely on optimization software such as Mosek or Gurobi to achieve this task for several penalty functions \citep{mosek, gurobi2019}.
	
	In the most general case where $p \notin \{1,2,\infty\}$, we utilize a simple algorithm relying on the \texttt{oem} package to estimate a penalized $L_p$ norm regression with an iteratively re-weighted least-squares algorithm like that described in Section 5.4 of  \cite{OsborneM.R.MichaelRobert1985Faio}.

	Additionally, we have developed additional software to perform best subsets, simulated annealing, and backward stepwise estimation of the desired loss functions to achieve models of the desired size. The algorithms to perform these tasks are detailed in \ref{sec:alg}.

	\subsection{SLIM-p estimation}
	In the case where we want to preserve an already existing linear model $m(z) = z \trans \hat{\theta} = \hat{\mu}_z$, we can formulate the $t^{\text{th}}$ sample of the interpretable prediction as in Eq. \ref{eq:SLIM-p} with
	$ \hat{\nu}_z^{(t)} = \sum_{j=1}^k z^{(j)} \hat{\theta}_j^{(t)} \hat{\alpha}_j$
	where $\hat{\alpha}_j \in \{0,1\}$ for an observation $z \in \mcB$. We will consider the values of $\hat{\theta}$ to be fixed in this setting and will only consider optimizing the 2-Wasserstein distance.
	Our problem is then to find the $\alpha$ and $\gamma$ that minimizes Eq. (\ref{eq:slimp_loss}).
	
	Ideally, we could use an optimization method to estimate both quantities jointly, but this equation is bilinear in its arguments and not guaranteed to be convex. Therefore, we need to break up the optimization problem into two steps: 
	\begin{enumerate}[label={Step \arabic{enumi}:}]
		\setlength{\itemindent}{.5in}
		\item Find the optimal transport solution $\gamma$ given $\alpha$ using Eq. (\ref{eq:OTfun}).
		\item Estimate a potential value for $\alpha$ given the current value of $\gamma$ using Eq. (\ref{eq:slimp_loss})
	\end{enumerate}
	\noindent For step one, we can use the algorithms in Section \ref{sec:ot_est} to estimate $\gamma$ while several methods are available for step two. We show that Eq. (\ref{eq:slimp_loss}) is a quadratic binary program so that we can utilize optimization techniques to find values of $\alpha$.
	
	\begin{proposition}
		Let $\boldz = \begin{pmatrix} z^{(1)} \otimes \hat{\theta}_1^{(1:T)} & \cdots & z^{(k)} \otimes \hat{\theta}_k^{(1:T)}
		\end{pmatrix} \trans $ be a $k \times T$ matrix, where $\otimes$ denotes the Kronecker product, and we have $N$ such matrices in the interpretable neighborhood $\mcB$.  We will denote the $t^\text{th}$ column of $\boldz$ as $\boldz^{(t)}  = \begin{pmatrix} z^{(1)} \hat{\theta}_1^{(t)} & \cdots & z^{(k)}\hat{\theta}_k^{(t)} \end{pmatrix}\trans$.
		Then Eq. (\ref{eq:slimp_loss}) is a quadratic program of the form 
		\[\alpha \trans \Sigma_{\boldz, \boldz}  \alpha - 2 \alpha \trans \Sigma_{\boldz, \hat{\mu}}, \]
		where 
		\begin{align}
			\Sigma_{\boldz, \boldz} &= \frac{1}{N\cdot T }   \sum_{\boldz \in \mcB}\boldz  \boldz \trans
			\label{eq:xtx}
			\intertext{and} 
			\label{eq:xty}
			\Sigma_{\boldz, \hat{\mu}} &= \frac{1}{N}   \sum_{\boldz \in \mcB}   \sum_{t =1,t' =1}^T \boldx^{(t')}\hat{\mu}_z^{(t)} \gamma_{t,t'} ,
		\end{align}
		and $\gamma$ is a transportation matrix.
		\label{prop:quad}
	\end{proposition}
	\noindent We defer the proof to  \ref{sec:quad}. 
	Equations (\ref{eq:xtx}) and (\ref{eq:xty}) will be useful because the set $\boldx_{1:N}$ could have dimensions as large as $k \times N \cdot T$, which will be quite large even for moderately sized problems, while the sufficient statistics will only have dimensions equal to the number of parameters $k$. Though not necessary, the additional scaling factor of $1/N$ in front of equations (\ref{eq:xtx}) and (\ref{eq:xty}) provides some stability computationally.
	
	This gives us at least two additional options to estimate $\alpha$. The solution to step two can be formulated as a quadratic binary programming problem:
	\begin{align}
		\label{eq:bin_prog}
		\hat{\alpha}_\lambda = \argmin_\alpha \,\, &  \alpha \trans \Sigma_{\boldz, \boldz}  \alpha - 2 \alpha \trans \Sigma_{\boldz, \hat{\mu}} \\
		\text{ such that } & \sum_{j=1}^k \alpha_j \leq \lambda,  \nonumber\\
		& \text{ for } \lambda \leq k \text{ and } \alpha \in \{0,1\}^k. \nonumber
	\end{align}
	Solvers are available to estimate $\alpha$ but the problem can be quite hard in general.
	
	Instead of solving the exact problem, we can find a relaxed solution for $\alpha$ in $\R^k$ using a sparsity inducing penalty, such as the $L_1$ norm, and then project that solution back to the feasible binary set. With this methodology, our unconstrained solution is
	\begin{equation}
		\tilde{\alpha}_\lambda = \argmin_{\alpha} \, \alpha \trans \Sigma_{\boldx, \boldx}  \alpha - 2 \alpha \trans \Sigma_{\boldx, \hat{\mu}} + P_\lambda(\alpha),
		\label{eq:alpha_unconst}
	\end{equation}
	where $P_\lambda(\cdot)$ is a convex penalty function like the LASSO, SCAD, or MCP \citep{Tibshirani1996, Fan2001, Zhang2010}. This solution is not guaranteed to be in the feasible set $\{0,1\}^k$, but we can move it back to the set of feasible solutions using the simple rule
	\begin{equation}
		\hat{\alpha}^{(j)}_\lambda = \left\{ \begin{array}{cl}
			1 &\text{ if } \hat{\alpha}_\lambda^{(j)} > 0.5,\\ 
			0 &\text{ otherwise.}  
		\end{array} \right. 
		\label{eq:selection}
	\end{equation}
	Essentially, we find the closest values in the feasible set to the unconstrained solution. Regardless of the method we choose to estimate $\alpha$, we can alternate steps one and two until convergence as detailed in Algorithm \ref{alg:slimp}. In practice, we find that the relaxed solution using a penalty function is much faster and performs well.
	
	\begin{algorithm}[t]
		\KwData{predictors for interpretation $z_{1:N}$ each in $\mcX \subseteq \R^k$, parameter estimates $\hat{\theta}^{(1:T)}$ each in $\R^k$, and original prediction $\hat{\mu}^{(1:T)}$ each in $\mcH^N$}
		\KwIn{penalty $\lambda$, tolerance $\epsilon$ }
		\KwResult{Sparse vector $\alpha \in \{0,1\}^k$}
		Set interpretable prediction $\hat{\nu}_i = z_i \trans \hat{\theta}^{(1:T)}, \, \forall i \in \{1,\ldots,n \}$\;
		Calculate the optimal transport matrix $\gamma = \text{OT}(\hat{\mu}, \hat{\nu})$ as in Eq. (\ref{eq:OTfun})\; 
		Calculate sufficient statistics $\Sigma_{\boldz, \boldz}$ and $\Sigma_{\boldz, \hat{\mu}}$ following Eqs. (\ref{eq:xtx}) and (\ref{eq:xty}), respectively\;
		Set $\hat{\alpha}_\lambda = 0$\;
		\While{ $\exists j \in \{1,\ldots,k\}: |\hat{\alpha}_\lambda^{(j)} - \alpha_j|/\alpha_j > \epsilon$}{
			Set $\alpha = \hat{\alpha}_\lambda$\;
			Calculate $\hat{\alpha}_\lambda$ as in Eq. (\ref{eq:bin_prog})  or as in Eqs. (\ref{eq:alpha_unconst}) and (\ref{eq:selection})\;
			Set $\hat{\nu}_i = \boldz_i \trans \hat{\alpha}_\lambda, \, \forall i \in \{1,\ldots,n\}$\;
			Calculate $\gamma = \text{OT}(\hat{\mu}, \hat{\nu})$ following Eq. \ref{eq:OTfun}\;
			Recalculate $\Sigma_{\boldz, \hat{\mu}}$  following Eq. (\ref{eq:xty})\;
		}
		\Return $\alpha$\;
		\caption{\textsc{quadraticSLIM-p}}
		\label{alg:slimp}
	\end{algorithm}
	
	In addition to this algorithm, we provide several other options to estimate $\alpha$ in  \ref{sec:alg} including a best subsets algorithm, a simulated annealing algorithm, and a backward stepwise algorithm.
	
	\section{Numerical experiments}\label{sec:sim}
	This section demonstrates the methods introduced in Section \ref{sec:setup} for a variety of data settings, sample sizes, and data dimensions. To solve the optimal transport problems in our SLIM-p methods, we utilize the exact, short simplex optimal transport method of \citet{Gottschlich2014}. 
	
	\subsection{Simulation setup}\label{sec:simsetup}
	\paragraph{Predictor Data} We test several conditions for our experiments to compare how correlation between predictor variables affect the results. In general, the predictor data is distributed as
	\[X_1, X_2,\ldots,X_n \iid \mcG\]
	with \[ \mcG \equivdist \N(0, \Omega)\]
	and \[\Omega\in \R^{k \times k} \]
	Each variable in $X_i$ has zero mean and unit-variance, but is correlated with the other variables in its group with correlation coefficient $\rho$ and uncorrelated with variables outside of its group. The variables are grouped in sets of 5. For example,  $X^{(1:5)}_i$ is the first group, $ X^{(6:10)}_i$ the second,  $X^{(11:15)}_i$, the third, and so on. This makes $\Omega$ a block-diagonal correlation matrix. We set $\rho$ to one of the values in $\{0, 0.5, 0.9\}$.
	\paragraph{Test data point} We generate a new test data point ${X}_0 \in \mcX$ from $\mcG$ as the data for which we wish to generate an interpretable model. Then we generate our neighborhood around ${X}_0$ as 
	\[Z_1,\ldots,Z_{N} \iid \N({X}_0, \hat{\Sigma}/n ),\]
	where $\hat{\Sigma} = \sum_{i=1}^n X_i X_i \trans$ and with $N = 3k$.
	\paragraph{Number of experiments}For each correlation, we run 100 experiments, unless otherwise noted
	
	With these simulation settings, we now describe a toy example to demonstrate how the method works.
	
	\subsection{Toy Example} \label{sec:toy}
	For our toy example, we run our experiment once with each of the three values of $\rho$ and a matrix of predictor data as described above. We set $n = 1024$ and $k = 5$. Then we sample an outcome from \[ Y_i \iid \N(X_i \trans \theta, \sigma^2)\] with $\theta =(-0.1, -0.2, 1.3, 1.4, 1.5)\trans$ and $\sigma^2=1$ and sample from a conjugate Normal posterior with priors
	\[\theta \sim \N_k(0, \ident_k), \] 
	and \[\sigma^2 \sim \InvGaT{1}{1}.\]
	
	We then pick the predictor value we wish to explain with an interpretable model as \[{x}_0 = (100,90,0.01,0.01,0.01)\trans.\] This is obviously a silly example since $ {x}_0^{(1:2)}$ are clearly outside the range of the data used to estimate the posterior distribution. However, it will be a useful test of our method, since we hope that the order of inclusion into the interpretable models is $2,1,5,4,3$, which would generate predictions close to the original model with the fewest covariates.
	
	\begin{figure}[!tb]
		\includegraphics[width=\textwidth]{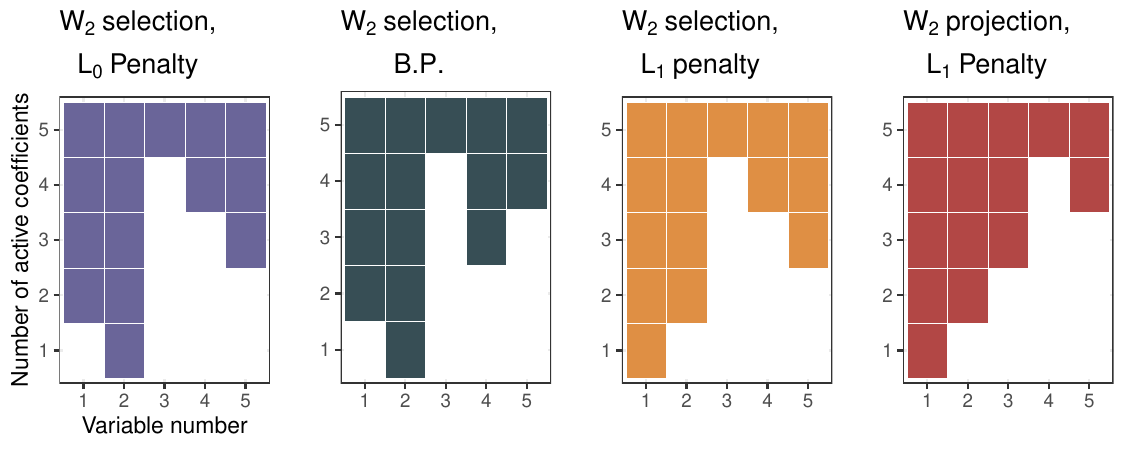}
		\caption
		{\textbf{Selection order for SLIM-p model with data correlation $\mathbf{\rho = 0.5}$.} We ran best subsets (first panel), a binary program (second panel), and relaxed quadratic binary program using an $L_1$ like penalty ( MCP net penalty with elastic net parameter 0.99 and MCP parameter 1.5) like in equations (\ref{eq:alpha_unconst}) and (\ref{eq:selection}) using a penalized regression (third panel), and for a $W_2$ projection with a the same MCP penalty (fourth panel). }
		\label{fig:slim_toy}
	\end{figure}
	\begin{figure}[!tb]
		\includegraphics[width=\textwidth]{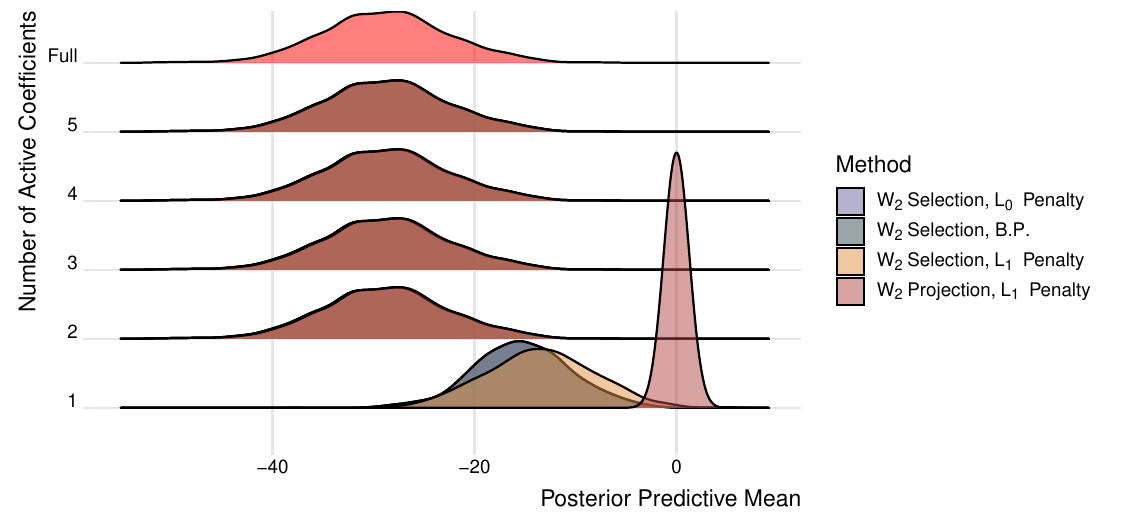}
		\caption
		{\textbf{Ridge plots for $\mathbf{\rho = 0.5}$.} The left panel is a ridge plot for the SLIM-a estimation and the right panel is a ridge plot for the SLIM-p estimation for the toy example in Section \ref{sec:toy}. }
		\label{fig:ridge_toy}
	\end{figure}
	
	We attempt both SLIM-a and SLIM-p estimation for this example. For SLIM-a estimation, we use the penalized regression described in Eq (\ref{eq:obj_l2}) with a neighborhood of points $Z_1,\ldots,Z_{100} \iid \N_k({x}_0, \hat{\Sigma}/n)$, where $\hat{\Sigma}$ is the empirical covariance of $X_{1:n}$. For SLIM-p estimations, we use an $L_0$ method, a penalized regression as in Eqs. (\ref{eq:alpha_unconst}) and (\ref{eq:selection}), a binary program like in Eq. (\ref{eq:bin_prog}), and a relaxed binary program like in Eq. (\ref{eq:alpha_unconst}) and (\ref{eq:selection}). We only display results for $\rho = 0.5$ but the results for $\rho \in \{0.0, 0.9\}$ are similar and can be found in  \ref{sec:extragraph}.
	
	In Figure \ref{fig:slim_toy}, we can see that the $L_0$ selects the same covariates in the order we would expect, while the relaxed binary program selects covariates in order of $1,2,5,4,3$. The binary program disagrees with the $L_0$ method about the covariate that should be third and fourth in the model. The $W_2$ projection method selects coefficients in a different order than the others. However, this does not affect predictions as we see in the next figure.
	
	The ridge plots in Figure \ref{fig:ridge_toy} also show that all methods approximate the full predictive distribution with two covariates, although the SLIM-a penalized regression does not find a good predictive model with only one covariate active.
	We can also see that the $L_0$ and binary program models do a slightly better job approximating the predictive distribution with for the first covariate.

	\subsection{Gaussian data example}
	For our Gaussian example we set $n = 1024$ and $k = 20$. We sample each $X_i$ as denoted above for $\rho \in \{0.0,0.5,0.9\}$, define a design matrix $\phi(X_i) = \begin{pmatrix} X_i& \left(X_{i}^{(1)}\right)^2 &  \left(X_{i}^{(3)}\right)^2 & \left(X_{i}^{(7)}\right)^2& X_{i}^{(13)} X_{i}^{(15)} \end{pmatrix}$, set $k'$ as the dimension of the design matrix ($k + 4$), and sample the outcomes as
	\[Y_i \sim \N(\phi(X_i) \trans \theta_{1:k'}  + \theta_0, \sigma^2).\]
	We set $\sigma^2 =1$ with $\theta$ as described below.
	
	\paragraph{Coefficient generation} For the various correlations of the predictor data, we scale our generated coefficients so as to give the linear predictor the same variance for easier comparison across conditions. 
	
	First, we draw one time from the following distributions and use the values for \emph{all} simulations:
	\begin{align*}
		\xi_0 &\sim \Un(1,2)\\
		\xi_{1:5} &\sim \Un(1,2)\\
		\xi_{6:10} &\sim \Un(-2,-1)\\
		\xi_{11:15} &\sim \Un(0,0.5)\\
		\xi_{16:20} &\sim \Un(-0.5,0)\\
		\xi_{21:24} &\sim \Un(0,0.5).
	\end{align*}
	Then we calculate our data-generating coefficients, $\theta$ utilizing a weighting function $w(\cdot, \cdot)$ that will take in the values of $\xi$ and the correlation matrix of the $\phi(X_i)$, $\Omega_\phi$, to appropriately scale predictions to have variance 1. We define $w: \R^{k'} \times \R^{k' \times k'} \mapsto \R^+ $ with 
	\[w(\xi_{1:(k + 4)}, \Omega_\phi) = \frac{1}{\sqrt{ \xi_{1:k}\trans \Omega_\phi \xi_{1:k} }}. \]
	From this, we set our $\theta$ vector as follows:
	\begin{align*}
		\theta_0 &= \xi_0\\
		\theta_{1:k} &= w(\xi_{1:k'}, \Omega_\phi) \xi_{1:k'} .
	\end{align*}
	For $\rho = 0,$ $\Omega_\phi$ will be the $k' \times k'$ identity matrix, $\ident_{k'}$. This means that $w(\xi_{1:k'}, \Omega) = 1/ \| \xi_{1:k'} \|_2$. Then the variance of the linear predictor is 
	
	\begin{align*}
		\var(\phi(x_i) \trans \theta_{1:k'} + \theta_0) &=  \frac{\xi_{1:k}\trans \var(\phi(x_i)) \xi_{1:k'}}{\| \xi_{1:k'} \|_2^2}\\
		&= \frac{\xi_{1:k'}\trans \ident_{k'} \xi_{1:k'}}{\| \xi_{1:k'} \|_2^2}\\
		&= \frac{\| \xi_{1:k'} \|_2^2}{\| \xi_{1:k'} \|_2^2}\\
		& = 1.
	\end{align*}
	
	\textbf{Posterior model.} We use a conjugate Gaussian model with priors $\sigma^2 \sim \InvGaT{1}{1}$ and $\theta_{0:k'} \iid \N(0,\sigma^2)$ and we then sample from the posterior $T=100$ times. 
	
	\paragraph{SLIM methods} Then we sample our test data and its neighborhood as described in Section \ref{sec:simsetup}. For this neighborhood, we run SLIM-a methods ($W_1, W_2, \text{ and } W_\infty$ penalized regressions), while on the single test data point we run two SLIM-p methods (binary program and relaxed relaxed binary program). For the penalized regression methods we utilize a MCP penalty with parameter equal to 1.1.
	
	\begin{figure}[!tb]
		\includegraphics[width=\textwidth]{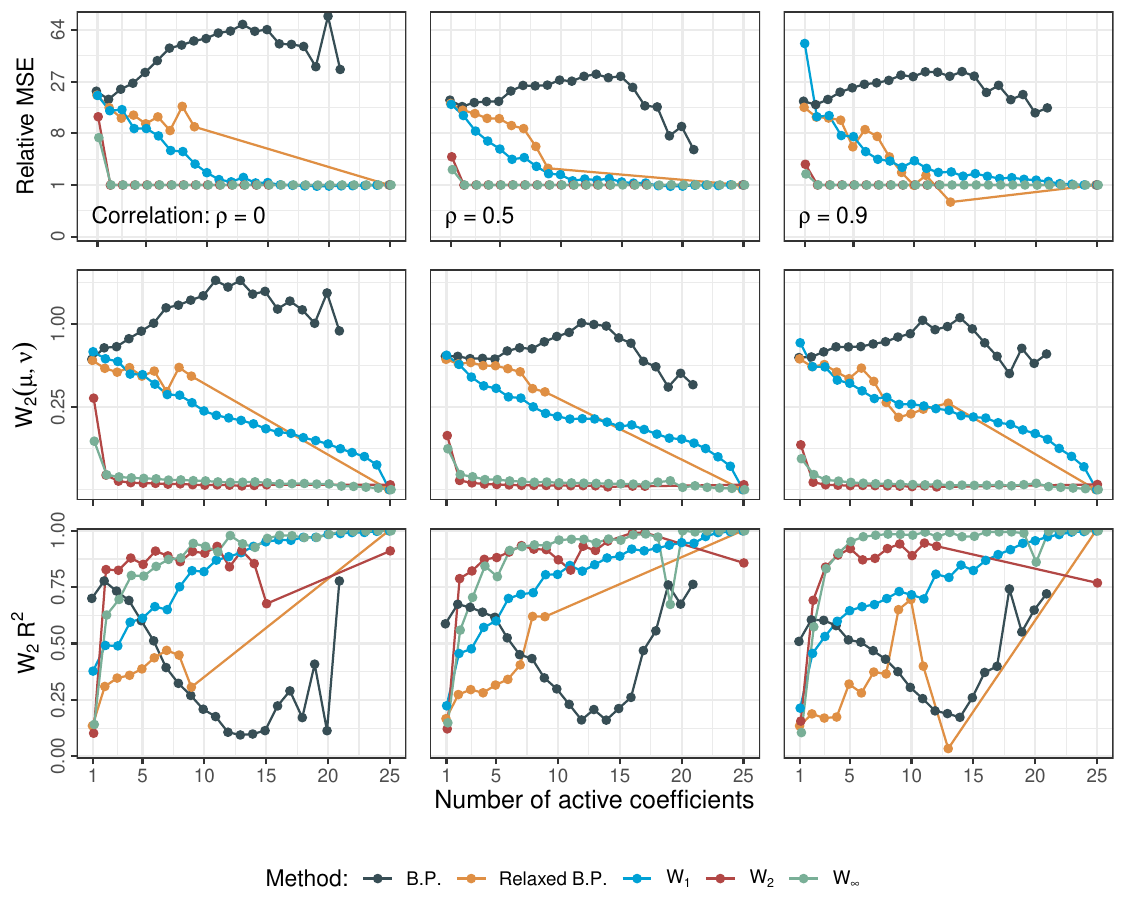}
		\caption
		{\textbf{Predictive evaluation on Gaussian outcome simulation with predictor correlations of, from left to right, $\mathbf{\rho = 0.0, 0.5, 0.9}$.} We evaluate the performance of our two SLIM-p models (binary program and relaxed binary program) and three SLIM-a models ($W_1$, $W_2$, and $W_\infty$ penalized regressions) in terms of relative mean-squared error (MSE), defined in Eq. (\ref{eq:relmse_1}), from the true mean parameter (top), 2-Wasserstein distance to the predictive distribution of the original model (middle) around our test point, and $W_2R^2$ values (bottom). 
			Figures represent averages across the 100 experiments.}
		\label{fig:gaussian_eg} 
	\end{figure}
	\begin{figure}[!tb]
		\includegraphics[width=\textwidth]{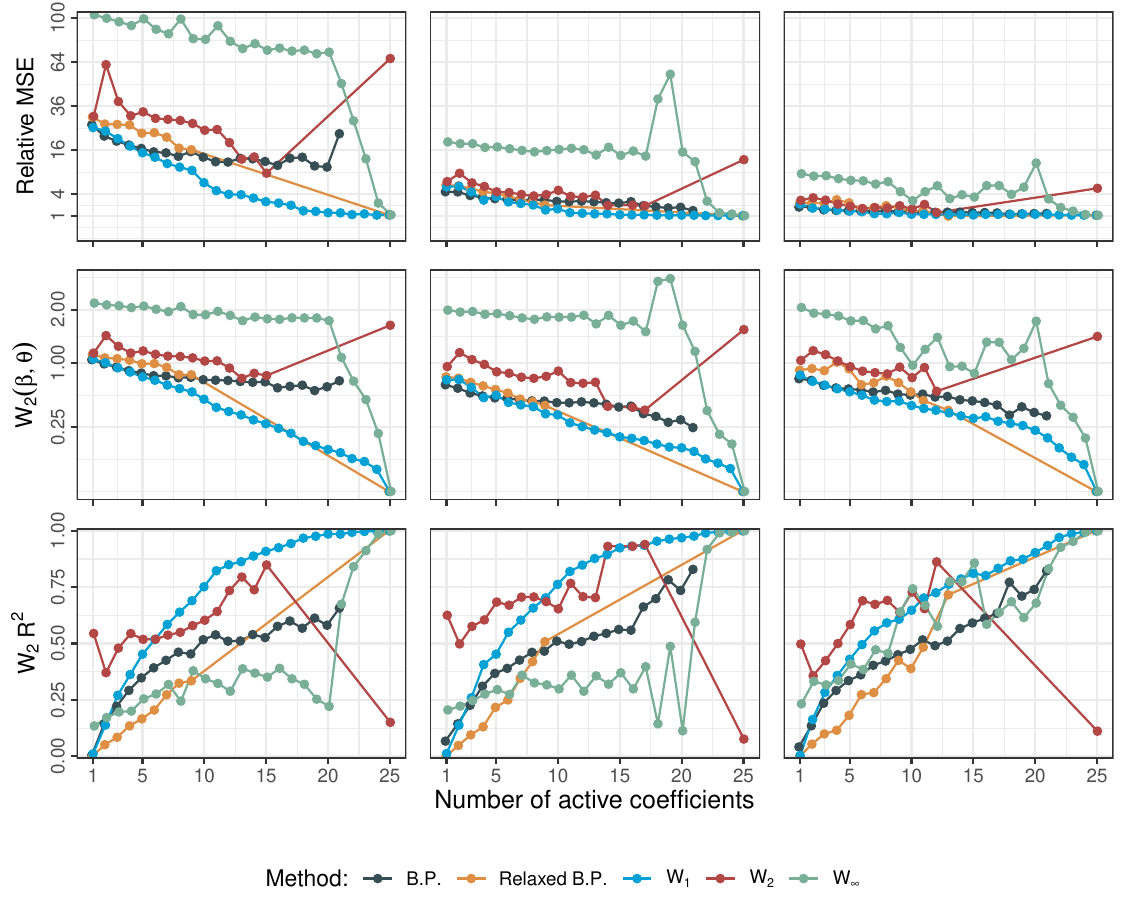}
		\caption
		{\textbf{Parameter evaluation on Gaussian outcome simulation with predictor correlations of, from left to right, $\mathbf{\rho = 0.0, 0.5, 0.9}$.} We evaluate the performance of our two SLIM-p models (binary program and relaxed binary program) and three SLIM-a models ($W_1$, $W_2$, and $W_\infty$ penalized regressions) in terms of relative mean-squared error (MSE), defined in Eq. (\ref{eq:relmse_2}), from the true coefficient vector (top), 2-Wasserstein distance to the posterior distribution of the full model coefficients (middle) around our test point, and $W_2R^2$ values from the corresponding Wasserstein distance (bottom). 
		Figures represent averages across the 100 experiments.}
		\label{fig:gaussian_eg_post} 
	\end{figure}
	
	\paragraph{Predictive loss evaluation} To evaluate the loss in predictive accuracy by using an interpretable model in an ideal world, we would examine the mean-squared error between the true mean parameter $\mu^\ast$  and our interpretable predictions relative to the mean-squared error of the original predictions and the true mean parameter. This quantity, which we call the relative mean-squared error, we define as 
	\begin{equation}
		\text{Relative MSE} \defn \frac{\E (\hat{\nu} - \mu^\ast )^2}{\E (\hat{\mu} - \mu^\ast )^2},
		\label{eq:relmse_1}
	\end{equation}
	where expectations are approximated with the empirical samples. Our goal is to then use this simulation setting to verify that the quantities we can evaluate in practice, the 2-Wasserstein distance between predictions and the corresponding $W_2R^2$, lead us to the same conclusions as when we have access to the true mean.
	
	Indeed, Figure \ref{fig:gaussian_eg} suggests that the 2-Wasserstein distance and $W_2R^2$ are suitable metrics to evaluate performance. For example, both the $W_2$ and $W_\infty$ regressions approximate the mean-squared error of the full model well with two covariates while the other models need more to do an adequate job. Similarly, we can see that both the $W_2$ and $W_\infty$ regressions approximate the full model distribution well after using two or three covariates while the $W_1$ regression needs more covariates to adequately approximate the full model.  Note that the graphs are the averages across all 100 experiments.
	
	\paragraph{Parameter loss evaluation} Since we are using the true model as our full model, we also evaluate how well the coefficients from the interpretable models approximate the true coefficients, $\theta$. Again, since we know the true parameter vector, we can construct a relative mean-squared error evaluating the closeness of the interpretable coefficients to the true parameters as compared to using the full model, defined in this case as
	\begin{equation}
		\text{Relative MSE} \defn \frac{\E \|\hat{\beta} - \theta \|_2^2}{\E \|\hat{\theta} - \theta \|_2^2},
		\label{eq:relmse_2}
	\end{equation}
	where expectations are again approximated with the corresponding empirical samples. Since these values will not be calculable in applied settings we again compare the relative mean-squared error to the 2-Wasserstein distance and the $W_2R^2$, this time between coefficient posteriors, in the hopes that similar conclusions can be drawn with these quantities.
	
	Even though both the SLIM-p methods do not perform as well as the SLIM-a methods on the predictive quantities, they are parameter preserving (see Figure \ref{fig:gaussian_eg_post}). Indeed, the binary program, relaxed binary program, and $W_1$ regressions better approximate the posterior on the coefficients. There appears to be a trade off between accurate approximations of the predictive or parameter distributions. Again, note that estimates are the averages across all 100 experiments.
	
	\subsection{Non-linear binary data example}
	\label{sec:bin_eg}
	For our second simulation example, we use a complicated non-linear data generating process to see how well the method can locally approximate a more complex response surface. We utilize the same data generation procedure for the predictors with $n = 131,072$ and $k=20$ with varying correlations of $\rho \in \{0.0, 0.5, 0.9\}$. However, we generate the outcome as
	\[ Y_i \sim \Bern{\text{logit}^{-1} \left [\mu_i\right]},\]
	with \[\mu_i =  h\left(x_i^{(1)}, x_i^{(2)}\right)  + 2 l\left(x_{i}^{(6)}\right) + r\left(x_{i}^{(7)}\right) + u \left(x_{i}^{(11)},x_{i}^{(15)} \right) ,\]
	where 
	\begin{align*}
		h(x, y)  &= \sin\left(\frac{\pi}{8}x y + \frac{\pi}{2} \right),\\
		l(x) &= \left \{ 
		\begin{array}{l} 
			1 \text{ if } x^2 > \frac{3}{8} \pi \\
			\cos\left(2 x^2  + \frac{\pi}{2}\right), \text{ otherwise}\end{array} 
		\right. , \\
		r(x) &= \left(\frac{2}{5} x^3 - \frac{1}{8} x^2 - 2 x\right)\exp\left(-\frac{x^2}{5}\right),\\
		\intertext{and}
		u(x,y) &=\tanh \left(-\frac{\pi}{8} xy^2 \right).
	\end{align*}
	\begin{figure}[!tb]
		\includegraphics[width=\textwidth]{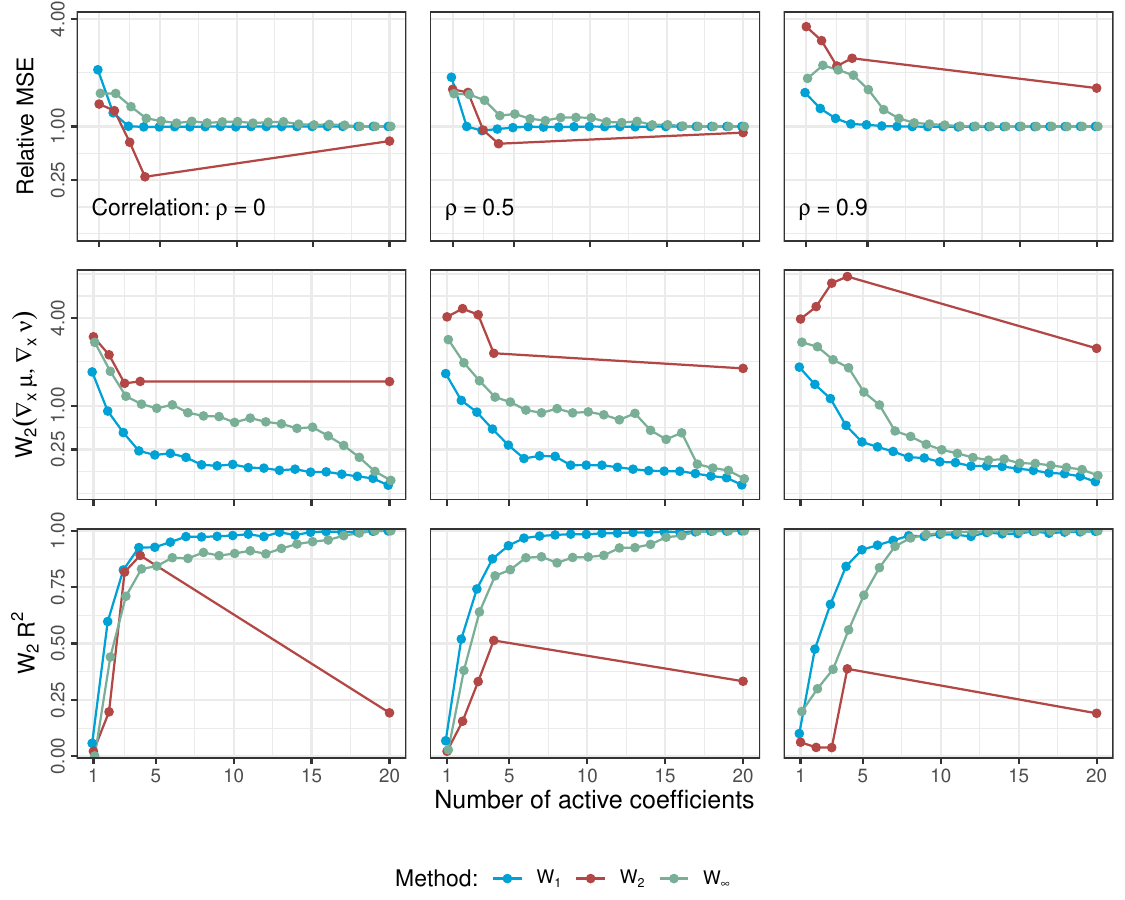}
		\caption
		{\textbf{Binary outcome simulation with correlations of, from left to right, $\mathbf{\rho = 0.0, 0.5, 0.9}$.} We evaluate the performance of our three SLIM-a models ($L_1$, $L_2$, and $L_\infty$ penalized regressions) in terms of how well they approximate the derivatives of the covariates at the evaluated points. The top chart plots the relative mean-squared error (MSE) of the interpretable and full models, defined in Eq. (\ref{eq:relmse_3}).
			The middle chart presents the 2-Wasserstein distance between the derivatives of the interpretable model (the coefficients) versus the derivatives of the covariates in the neural network and the bottom prents the corresponding $W_2R^2$.  We can see that both the $W_1$ and $W_\infty$ regression approximate the derivatives of the neural network well with 3 to four coefficients, while for smaller correlations the $W_2$ model actually does a better job of approximating the true derivaives than the derivatives of the neural network. Presented figures represent averages across the 100 experiments.}
		\label{fig:binomial_eg}
	\end{figure}
	
	\paragraph{Neural network model} We estimate a 4-layer, feed-forward neural network with ELU activating functions for the full model. The first hidden layer has 200 nodes and each subsequent layer before the last has 14. To estimate the uncertainty of our predictions, we use $T=100$ bootstrap samples. The model was constructed in Python 3.7 using the PyTorch package and interfaced with R using the \texttt{reticulate} package \citep{10.5555/1593511, NEURIPS2019_9015, Ushey2020}.

	\paragraph{SLIM methods} Then we sample our test data and its neighborhood as described in Section \ref{sec:simsetup} in order to approximate the derivatives with respect to the coefficient in the neural network. For this neighborhood, we run SLIM-a methods ($W_1, W_2, \text{ and } W_\infty$ penalized regressions) again with an MCP penalty and parameter set to 1.1. We don't utilize SLIM-p methods since there is no ``true'' linear model of interest.
	
	\paragraph{Evaluation} Since the Gaussian neighborhood is so small given our large sample size because $Z_{1:N} \iid \N(X_0, \hat{\Sigma}/n)$, we change our evaluative metric slightly. Instead of measuring predictive performance, we evaluate how well an interpretable model fit on this small neighborhood approximates the derivatives of the true function with respect to the input covariates. This example will also give a sense of the range of interpretable possibilities of the SLIM methods. First, we compare the mean-squared error of derivatves from the interpretable models (the coefficients from the linear models) to the true derivatives of the data generating function relative to the mean-squared error of the derivatives of the neural network to the true data generating functions, defined as
	\begin{equation}
		\text{Relative MSE} \defn \frac{\E \|\nabla_x \hat{\nu} - \nabla_x \mu \|_2^2}{\E \|\nabla_x \hat{\mu} - \nabla_x \mu \|_2^2} = \frac{\E \|\hat{\beta} - \nabla_x \mu \|_2^2}{\E \|\nabla_x \hat{\mu} - \nabla_x \mu \|_2^2}.
		\label{eq:relmse_3}
	\end{equation}
	Again we utilize the 2-Wasserstein distances and $W_2R^2$ to evaluate the practical performance 
	of the interpreatable models in the hopes that they approximate the case when we know the truth.
	
	In Figure \ref{fig:binomial_eg}, we see that there is some discrepancy between the mean-squared error evaluation and the other quantities for correlation 0 and 0.5 for the $W_2$ projections. However, for the $W_1$ and $W_\infty$ models, the MSE and Wasserstein quantities agree.
	
	\section{Applied data analysis}
	In this section, we apply two different types of models to two different cancer datasets to demonstrate the utility of the SLIM method. First, we utilize a Cox proportional hazards model \citep{Cox1972} to estimate the recurrence of ovarian cancer and demonstrate how the methodology can be used to approximate the estimate and distribution of the partial maximum likelihood estimator (MLE). Second, we demonstrate the method in a Bayesian additive regression tree model that predicts survival from glioblastoma \citep{Chipman2010}. 
	
	\begin{figure}[!b]
		\centering
		\includegraphics[width = \textwidth]{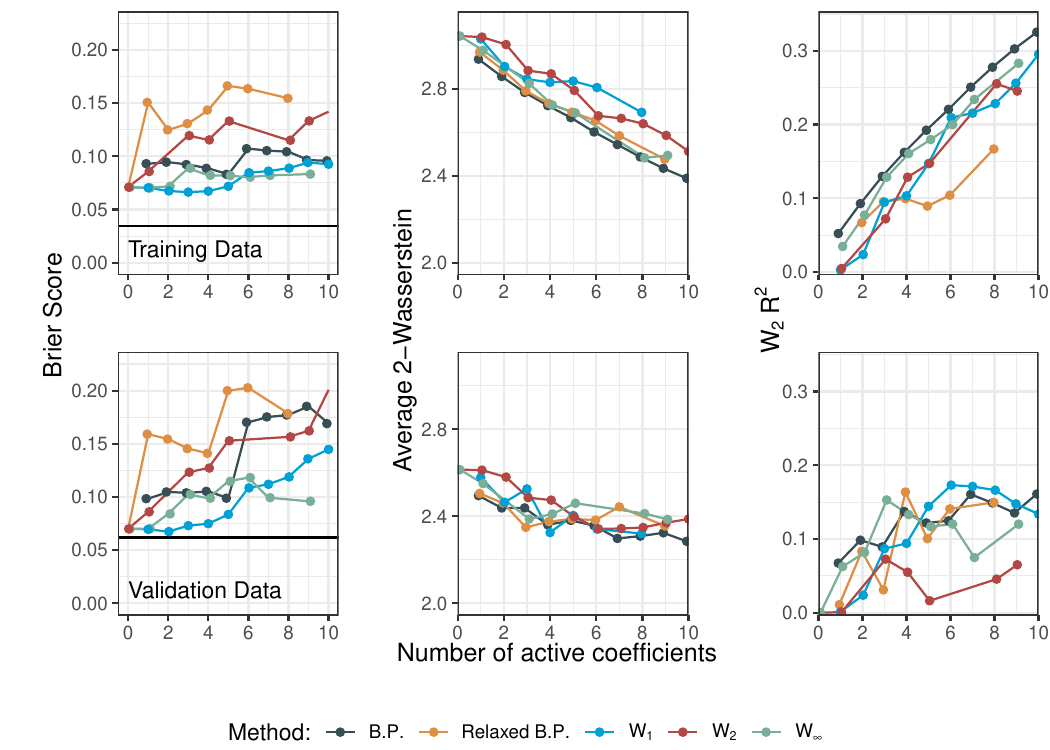}
		\caption{\textbf{Performance of training (top) and validation (bottom) data in the curated ovarian data.} The left most column displays the Brier score of the interpretable models as compared to the average Brier score of the original model (solid line). The center column displays the average 2-Wasserstein distance from the full distribution and the right column displays the $W_2R^2$ with the ``null'' distribution being the average value of the linear predictor across the entire data set giving  $W_2R^2$ the interpretation analogous to traditional $R^2$.}
		\label{fig:ovar_train}
	\end{figure}
	
	\subsection{Curated ovarian data}
	For our first applied analysis, we utilized participant data from \cite{Bell2011} as collected by the \texttt{R} package \texttt{curatedOvarianData} \citep{Ganzfried2013}. Cleaning the data and removing observations with missing data leaves 543 individuals. Then we selected covariates through leave-one-out cross validation using a Cox regression with a Lasso penalty as implemeted by the \texttt{glmnet} package in R \citep{Friedman2010}. We chose the covariates active in the model with the minimum cross-validation error---leaving 130 covariates for the final model. Finally, we split the data into a training set ($n$ = 489), a validation set ($n = 53$), a test point ($n=1$). The data as used is available from the \texttt{R} package found on GitHub at \url{https://www.github.com/ericdunipace/SLIMpaper}.
	
	
	
	Our predictive model in this case is a maximum likelihood Cox proportional hazards model and checks on the validation data indicate the model performs reasonable well. For example, on the validation data the concordance of the model is 0.7579 and the integrated Brier score is 0.0571 \citep{BRIER1950}. After these checks, we sampled 100 times from the asymptotic distribution of the model. Based on these samples from the asymptotic distribution, the average integrated Brier score is 0.03453 in the training data and 0.06202 in the validation data.
	
	Next, we ran both global and local versions of our SLIM models and evaluated their performance using a variety of metrics. For the global version, we estimated our SLIM models on the training data and then measured their performance on both the training and validation data. We can see in the Figure \ref{fig:ovar_train} that the models perform a little bit worse in the validation data but that in both cases we might conclude that the binary program performs best for a given model size (middle and right columns). However, in terms of Brier score (left column)
	the performance of the binary program is more mixed and it appears that relaxed binary program performs better. However, all methods are relatively close in terms of prediction. Finally, we detail the best, median, and worst case predictions in terms of their ridge plots in Section \ref{sec:data_graph}. Note that the models with zero coefficients only include the estimated baseline hazard function from the MLE model. As such, they can be interpreted as a pseudo-intercept only model.
	\begin{figure}[!tb]
		\centering
		\includegraphics[width=\textwidth]{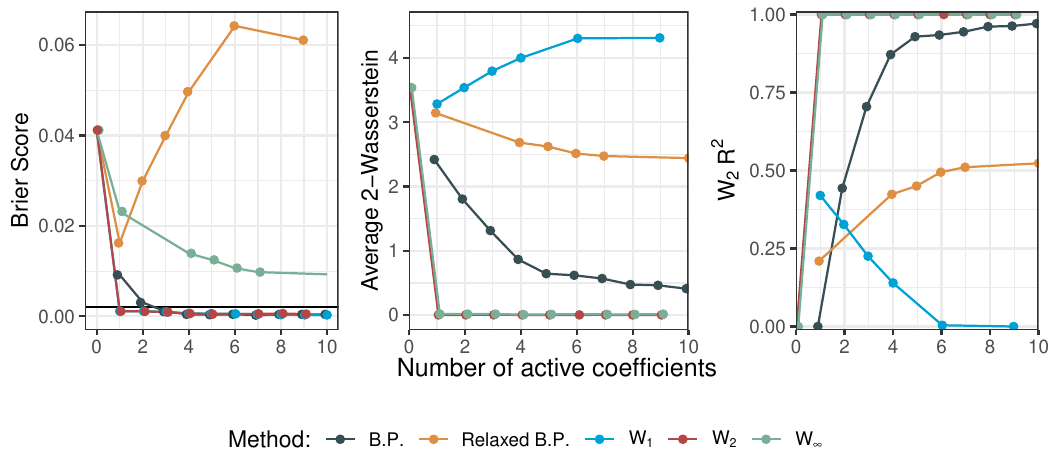}
		\caption{\textbf{Performance of test data in the curated ovarian data.} The left most column displays the Brier score of the interpretable models as compared to the average Brier score of the original model. The center column displays the average 2-Wasserstein distance from the full distribution and the right column displays the $W_2R^2$.
		}
		\label{fig:ovar_local}
	\end{figure}
	
	For the local version, we constructed SLIM-a/p using our test data point, and its corresponding neighborhood $\mcB$, and evaluated the interpretable models' performance on the test data point. Of the SLIM-p models, we can see that the binary program model appears to do fairly well while the relaxed binary program has  difficulty recovering the original prediction. In the SLIM-a models, the $W_2$ and $W_\infty$ models both perform well across all of metrics but the $W_1$ model generates a reasonable Brier score even though the 2-Wasserstein distance and $W_2R^2$ do not look satisfactory.
	
	\subsection{Glioblastoma data}\label{sec:gbm_sec}
	Our second data set concerns a sample of confidential glioblastoma patients from the Dana Farber Cancer center. After cleaning and processing the data files we have 157 patients with 6 covariates including age, gender (male or female), methylation status (yes or no), resection (biopsy, partial, or total), and a binary variable equal to 1 if the patient's Karnovsky prognostic score is above 80 and 0 otherwise. We then partitioned our sample into a training set of 141 patients, a validation set of 15 patients, and a test set of 1 patient.
	
	\begin{figure}[!tb]
		\centering
		\includegraphics[width =\textwidth]{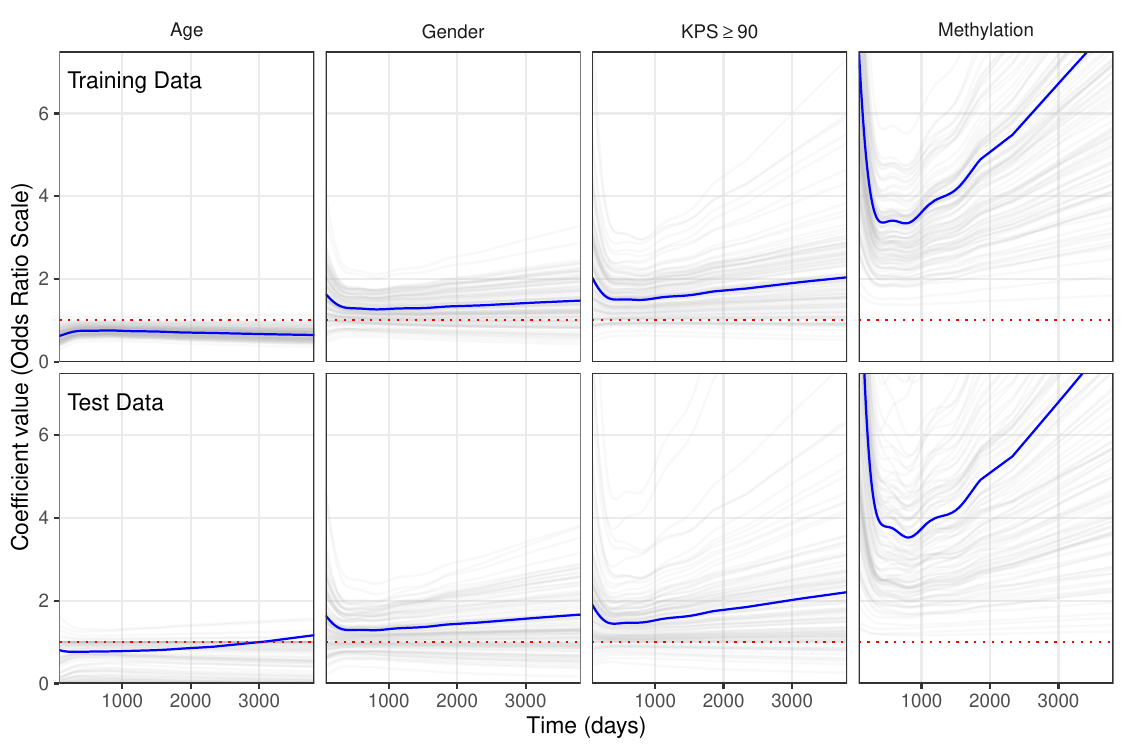}
		\caption{\textbf{Coefficient effect over time in the interpretable GAM model for the GBM data.} The effects in the figures are on the odds ratio scale where a single unit change in the coefficient multiplies the odds of survival by the vivsualized amount. Blue lines denote the average trajectory and in gray are 100 trajectories from the set of fitted GAM models.}
		\label{fig:gbm_or}
	\end{figure}
	
	\begin{table}[!bt]
		\centering
		\begin{tabular}{rrrr}
			\hline
			& Training data & Validation data & Test data \\ 
			\hline
			$W_2 R^2$ & 0.85 & 0.82 & 0.82 \\ 
			\hline
		\end{tabular}
		\caption{\textbf{$W_2 R^2$ statistics for the the training, validation, and test data for the Glioblastoma data set.}
			The null models in each case are an intercept only model using the resection status
			over time. The models are the same in the training and validation cases meaning that
			the validation data evaluates an out of sample fit. The test data model was
			fit on an interpretable neighborhood around the test point and then evaluated
			at the test data.} 
		\label{tab:w2r2_gbm}
	\end{table}
	
	The outcome model was an accelerated failure time Bayesian additive regression tree (BART) model to predict survival in the training sample \citep{Chipman2010, McCulloch2019}. Our model uses 50 trees in the summation with base parameter set to 0.95 and the power parameter set to 0.25. We ran the sampler for 200,000 warm-up iterations and then sampled 200,000 times from the posterior. However, we only kept every 100th sample to reduce the memory requirement of storing the final model, giving a total of 2000 samples. We evaluated the performance of the model on the validation set by calculating an integrated Brier score and examining the Brier score over time. Results are presented in \ref{sec:data_graph}.
	\begin{figure}[!ptb]
		\centering
		\includegraphics[width=\textwidth]{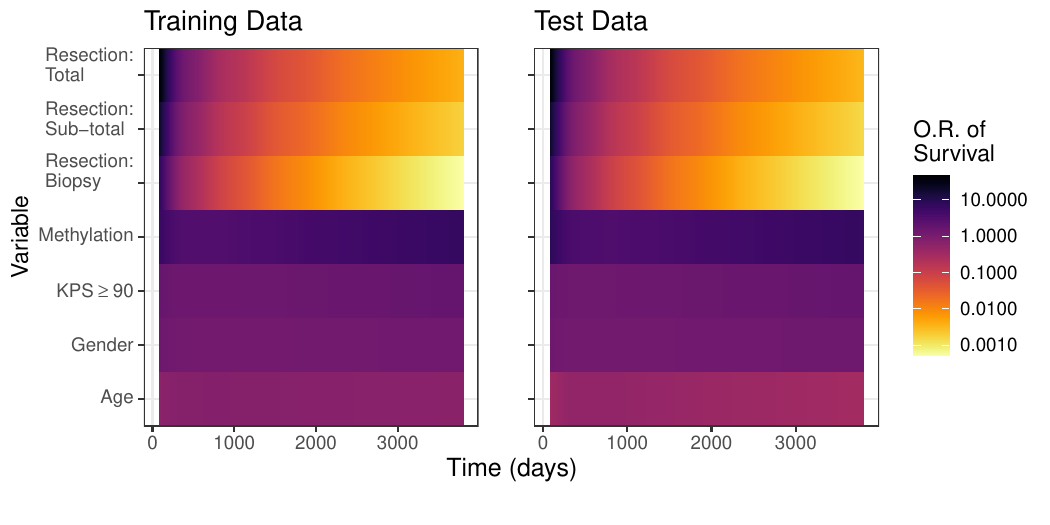}
		\caption{The coefficient effect in terms of odds ratio presented for the interpretable GAM models on the GBM data. Dark colors indicate a covariate effects survival positively at a given time while bright yellow colors indicate that a covariate negatively impacts survival at a given time. Graphs presented for training data and the test data point.}
		\label{fig:gbm_heat}
		
		\vspace*{\floatsep}
		
		\includegraphics[width=\textwidth]{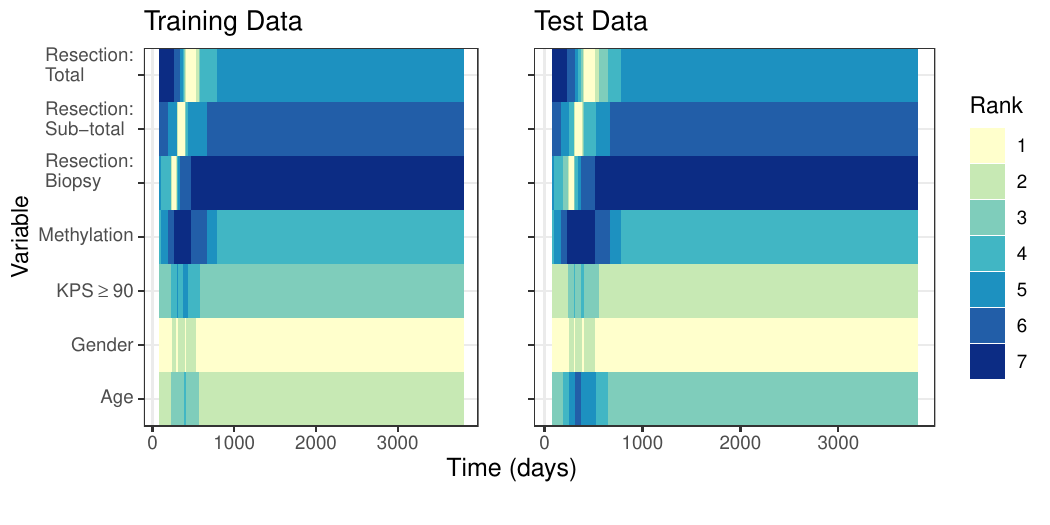}
		\caption{Coefficients ranked by their absolute value over time for the interpretable GAM models on the GBM data. Dark colors indicate the covariate had the largest absolute coefficient at that time indicating it was potentially most impactful to survival. Lighter colors indicate smaller absolute coefficient values, indicating that these coefficients were potentially less important. Graphs presented for training data and the test data point.}
		\label{fig:gbm_rank}
	\end{figure}
	
	Then we developed our interpretable models as follows. First, for the training, validation, and test data we predicted survival curves for each observation in the respective data and transformed them to the logit scale. Then to each sample we fit a generalized additive model (GAM) to allow coefficients to vary over time but to retain an interpretability on the coefficients. For the training data, the data matrix used to estimate the original model was used. For the test data, a data matrix using a neighborhood around the individual's age and every combination of the discrete covariates (gender, KPS, methylation, and resection status) was used. Finally, we evaluated the performance of each model by calculating a $W_2R^2$ statistic to determine how well the models approximate the data relative to the worst performing model for a given method (table \ref{tab:w2r2_gbm}).
	
	To ease interpretation of the models, we plotted the coefficients not related to resection status over time to see how the effects change for both the models fit on the training and the test data (Figure \ref{fig:gbm_or}). Overall, the fits between the training and test data are similar but with two noticeable differences: the test data model is more variable and the effect of age on the test data also changes on average from negative to positive towards the end of the follow-up period. However, the ultimate effects of these covariates are dwarfed by the effect of resection status. Ultimately, everyone dies from GBM and the heat maps of coefficient effects in Figures \ref{fig:gbm_heat} and \ref{fig:gbm_rank} show this.
	
	\section{Discussion}
	In this paper we have shown the utility of Wasserstein distances for model interpretation and provided a variety of computational tools to estimate and evaluate these quantities. The work of estimating models is hard but making them interpretable is just as challenging. Our hope is that we have made this task a bit easier for the applied researcher. We also note that that are several potential future directions for this methodology. 
	
	First, this work can be extended by adding in tools from the generalized estimating equation (GEE) literature. This would allow the estimation of models on the scale of the mean of the outcome. The only challenge would be to find an efficient way of estimating these models for multivariate data with group penalties.
	
	Second, the method is likely directly applicable to model reduction in the style of \citet{Goutis1998}
	and \citet{Dupuis1998}, but without likelihood assumptions. We expect that this methodology might be of interest in model reduction for the approximate Bayesian computation literature since one does not need to assume a functional form on the model probability distribution. This idea warrants more examination.
	
	In conclusion, Wasserstein distances have great utility in many fields. One such field is in the realm of model interpretation and our results show how helpful the concepts of optimal transport can be when applied to any arbitrarily complicated model.

	\clearpage

	
	\section*{Acknowledgements}
		We would like to thank Claire Chaumont for helpful comments
		and feedback
		
	\section*{Funding}
	This research was funded by generous support from NIH grant 5T32CA009337-40 and the Department of Biostatistics at the Harvard T.H. Chan School of Public Health
	
	\bibliographystyle{plainnat}
	\bibliography{slim}
	\newpage
	
	\appendix
	\section{Additional algorithms}\label{sec:alg}
	In addition to the algorithms detailed in Section \ref{sec:comp}, we also developed simulated annealing, backwards stepwise, and Hahn-Carvalho-based (HC) methods for SLIM-a and SLIM-p estimation. The SLIM-a and SLIM-p algorithms only differ in how they calculate the coefficients for the interpretable model. As a reminder, the interpretable model is of the form $\sum_{i = 1}^n x_i \trans \beta$ and for SLIM-a, $\beta$ is estimated via a penalized linear regression while for SLIM-p, $\beta = \diag(\alpha) \hat{\theta}$ with $\alpha$ being an indicator if a dimension of $\hat{\theta}$ is on or off. For more details please see Sections \ref{sec:slima} and \ref{sec:slimp}. We denote the general function to calculate coefficients \textsc{InterpretableCoefficient} and refer to the SLIM-a flavor as \textsc{Adaptive} and the SLIM-p flavor as \textsc{Fixed} which we detail in algorithms \ref{alg:adaptive} and \ref{alg:fixed}, respectively. 
	
	If we are doing a SLIM-a estimation, then we run full $L_p$ projection of $\hat{\mu}$ into $x^{(\mcA)}$ by minimizing $ \sum_{i=1}^n  \| \hat{\mu}_i - x_i^{(\mcA)} \beta \|^p_p$ over $\beta$. 
	If we are doing a SLIM-p estimation, the we run an $L_2$ regression and return a sparse version of $\hat{\theta}$ with dimensions set to 0 if they are not in the active set $\mcA$.
	Additionally, the \textsc{Adaptive} function ignores the values of $\hat{\theta}$ while the \textsc{Fixed} function ignores the values of $\hat{\mu}$; however, both functions take in these unnecessary values since the general function  \textsc{InterpretableCoefficient} needs one or the other.
	
	\begin{algorithm}[H]
		\KwData{predictor data for interpretation $x_{1:N} \in \mcX^{k}$, posterior samples $\hat{\theta}^{(1:T)}$ each in $\Theta$, active set $\mcA \subseteq \{1,\ldots,k\}$, original prediction $\hat{\mu}$ in $\mcH^{N \times T}$}
		\KwResult{sparse coefficient matrix $\hat{\beta} \in \R^{k \times T}$}
		Set $\tilde{\beta} = \argmin_\beta \sum_{i=1}^N ({\mu}_i - \sum_{j \in \mcA }x_i^{(j)} \beta_j)^2$\;
		$ {\beta}_j = \begin{cases}
			\tilde{\beta} \text{ if } j \in \mcA \\ 
			0 \text{ if } j \notin \mcA
		\end{cases} $\;
		\Return $\hat{\beta}$\;
		\caption{\textsc{Adaptive}$(x_{1:N}, \theta, \mcA, \mu)$ }
		\label{alg:adaptive}
	\end{algorithm}
	
	\begin{algorithm}[H]
		\KwData{predictor data for interpretaion $x_{1:N} \in \mcX^{k}$, posterior samples $\hat{\theta}^{(1:T)}$ each in $\Theta$, active set $\mcA \subseteq \{1,\ldots,k\}$, original prediction $\hat{\mu}$ in $\mcH^{N \times T}$}
		\KwResult{sparse coefficient matrix $\hat{\beta} \in \R^{k \times T}$}
		$ {\beta}_j = \begin{cases}
			\hat{\theta}_j \text{ if } j \in \mcA \\ 
			0 \text{ if } j \notin \mcA
		\end{cases} $\;
		\Return $\hat{\beta}$\;
		\caption{\textsc{Fixed}$(x_{1:N}, \theta, \mcA, {\mu})$ }
		\label{alg:fixed}
	\end{algorithm}
	
	\subsection{Best subsets algorithm}\label{sec:l0}
	One method might be to check all possible subsets of covariates for a given model size. This leads to an $L_0$ penalty function, effectively checking all possible combinations of $\{1,\ldots,k\}$ for a given model size. However, this task quickly becomes computationally infeasible after approximately $k=20$ variables since there would be $2^k$ possible parameter combinations to check for linear models. However, this is feasible for small models and we detail this method in algorithm \ref{alg:l0}.
	
	\begin{algorithm}
		\KwData{predictor data for interpretation $x_{1:N}$ each in $\mcX \subseteq \R^k$, $\hat{\theta}^{(1:T)}$ each in $\Theta$, and $ \hat{\mu}^{(1:T)}$ each in $\mcH^N$}
		\KwResult{ Set of matrices  $B_{1:(k-1)}$ of sparse coefficients for a given model size  }
		\For{$j=1$ \KwTo $k-1$ }{
			Set $\mcS = \{\mcA \subset \{1,\ldots,k\}: |\mcA| = j\}$, the set of all possible subsets of size $j$\;
			\For{$s = 1$ \KwTo  $|\mcS|$}{
				Set $\mcA= \mcS_s$\;
				Set $\hat{\beta} = \operatorname{\textsc{InterpretableCoefficient}}(x_{1:N}, {\theta}, \mcA, {\mu})$\;
				Set $\hat{\nu}_i =  \sum_{j  = 1}^k x_i^{(j)} \hat{\beta}_j$, $\forall i \in \{1,\ldots,N\}$\;
				Set $\boldw_s =\approxwass{\mu}{\nu}$\;
			}
			Set $ l = \argmin_{s } \{\boldw_s < \boldw_{s'}, \forall s' \neq s\}$\;
			Set $B_{j} = \operatorname{\textsc{InterpretableCoefficient}}(x_{1:N}, {\theta}, \mcS_l, \mu)$\;
		}
		\Return  $B_{1:(k-1)}$\;
		\caption{\textsc{BestSubsets}$(x_{1:N}, \theta, \mu)$ }
		\label{alg:l0}
	\end{algorithm}
	
	\subsection{Simulated annealing algorithm}\label{sec:sa}
	The general idea behind the simulated annealing algorithms detailed below is that for a given model size, the algorithm will propose removing one coefficient from the interpretable model and adding one coefficient the model. The proposal is then accepted with a Metropolis-Hastings step with relying on a temperature sensitive Boltzman probability distribution \citep{Hastings1970}.  With an appropriate proposal function, this algorithm can be an efficient exploration of the interpretable model space. Without an inappropriate proposal function, the exploration of the model space will be a random walk that may not adequately explore the function minima. 
	
	\begin{algorithm}
		\KwData{predictor data for interpretation $x_{1:N} \in \mcX^{N}$, posterior samples $\hat{\theta}^{(1:T)} \in \R^{k \times T}$, and original prediction $\hat{\mu}^{(1:T)} \in \mcM^{N \times T}$}
		\KwIn{vector of temperatures $R$, total iterations $L$, desired model size $k'\in \Nat$, and proposal function $\boldK$ for swapping states }
		\KwResult{sparse coefficient matrix $\hat{\beta} \in \R^{k \times T}$}
		Initialize the active set $\mcA = \{j \in \{1,\ldots,k\}: |\mcA| = k' \}$ uniformly at random\;
		Initialize the non active set $\mcC = \{1,\ldots,k\} \setminus \mcA$\;
		Set $\hat{\beta} = \operatorname{\textsc{InterpretableCoefficient}}(x_{1:N}, \theta, \mcA, {\mu})$\;
		Set ${\nu}_i =  \sum_{j  = 1}^k x_i^{(j)} \tilde{\beta}_j$, $\forall i \in \{1,\ldots,N\}$\;
		Calculate initial $p$-Wasserstein distance for active set, $\boldw = \wass{{\mu}}{{\nu}}{p}$\;
		\ForEach{$r \in R$} {
			Set $l = 0$;
			\While{$l < L$} {
				$l = l + 1$\;
				Propose new active and non-active sets $\{\tilde{\mcA},\tilde{\mcC}\} = \boldK(\{\mcA, \mcC \}\mapsto \cdot)$\;
				Set $\tilde{\beta} = \operatorname{\textsc{InterpretableCoefficient}}(x_{1:N}, {\theta},\tilde{\mcA}, {\mu})$\;
				Set ${\nu}_i =  \sum_{j  = 1}^k x_i^{(j)} \tilde{\beta}_j$, $\forall i \in \{1,\ldots,N\}$\;
				Calculate new approximate or exact 2-Wasserstein distance $\tilde{\boldw} =\approxwass{{\mu}}{{\nu}}$\;
				$U \sim \Un(0,1)$\;
				\If{$U < \min\left(1, \frac{\exp(-\tilde{\boldw}/r)}{\exp(-\boldw/r)}  \frac{K(\{\tilde{\mcA}, \tilde{\mcC} \} \mapsto \{\mcA, \mcC \})}{K(\{\mcA, \mcC \} \mapsto \{\tilde{\mcA}, \tilde{\mcC}\})} \right)$}{
					$\mcA = \tilde{\mcA}$\;
					$\mcC = \tilde{\mcC}$\;
					$\boldw = \tilde{\boldw}$\;
					$\hat{\beta} = \tilde{\beta} $\;
				}
			}
		}
		\Return $\hat{\beta}$
		\caption{\textsc{SimulatedAnnealing}$(x_{1:N}, \theta, \mu, R, L, k', \boldK)$ }
		\label{alg:simanneal}
	\end{algorithm}
	
	\subsection{Backward stepwise algorithm}\label{sec.sw}
	The backward stepwise algorithm starts with all covariates included in the model. Then at each step it removes the covariate that leads to the smallest change in 2-Wasserstein distance to the original model. 
	Another way of thinking about this algorithm is that at each step it removes the least important variable from the current active set, conditional on the currently active set. As in the simulated annealing algorithms detailed in Section \ref{sec:sa}, the SLIM-a (algorithm \ref{alg:adaptive}) and SLIM-p (algorithm  \ref{alg:fixed}) methods differ only in how they estimate $\hat{\nu}$.
	
	As opposed to an $L_0$ approach, this method would have to check $\sum_{j=1}^k j = \frac{1}{2} k(k+1) $ models rather than $2^k$. However, the stepwise approach has the problem that it may not select the variables that minimize the 2-Wasserstein distance for a given model size. 
	
	\begin{algorithm}
		\KwData{predictor data to interpret $x_{1:N}$ each in $\mcX \subseteq \R^k$, $\hat{\theta}^{(1:T)}$ each in $\Theta$, and $ \hat{\mu}^{(1:T)}$ each in $\mcH^N$}
		\KwResult{ Set of matrices  $B_{1:(k-1)}$ of sparse coefficients for a given model size  }
		\For{$j=1$ \KwTo $k-1$ }{
			Set $\mcA^{(j)} = \{s: \alpha_s \neq 0\}$\;
			\ForEach{$s \in \mcA^{(j)} $}{
				Set $\tilde{\mcA} = \mcA^{(j)} \setminus s$\;
				Set $\hat{\beta} = \operatorname{\textsc{InterpretableCoefficient}}(x_{1:N}, {\theta}, \tilde{\mcA}, {\mu})$\;
				Set ${\nu}_i =  \sum_{j  = 1}^k x_i^{(j)} \hat{\beta}_j$, $\forall i \in \{1,\ldots,N\}$\;
				Set $\boldw_s = \approxwass{{\mu}}{{\nu}}$\;
			}
			Set $ l = \argmin_{s \in \mcA^{(j)}} \{\boldw_s < \boldw_{s'}, \forall s' \in  \mcA^{(j)} \setminus s\}$\;
			Set $B_{j} = \operatorname{\textsc{InterpretableCoefficient}}(x_{1:N}, {\theta}, \tilde{\mcA}, {\mu})$\;
		}
		\Return  $B_{1:(k-1)}$\;
		\caption{\textsc{Stepwise}$(x_{1:N}, \theta, \mu)$ }
		\label{alg:step}
	\end{algorithm}
	
	%
	
	\clearpage
	
	\section{Proof of proposition \ref{prop:quad}} \label{sec:quad}
	\begin{pf}
		For $t \in \{1,\ldots,T\}$ and $i \in \{1,\ldots,N\}$, we can write
		\[ 
		\hat{\nu}_i^{(t)}(\alpha) = \sum_{j=1}^k x_{i}^{(j)} \hat{\theta}_j^{(t)} \alpha_j = \begin{pmatrix} x_{i}^{(1)} \hat{\theta}_1^{(t)} & \cdots & x_{i}^{(k)} \hat{\theta}_k^{(t)} \end{pmatrix} \alpha,
		\] 
		and
		\[
		\hat{\nu}_i^{(1:T)}(\alpha) =  \begin{pmatrix}
			x_{i}^{(1)} \otimes \hat{\theta}_1^{(1:T)} & \cdots & x_{i}^{(k)} \otimes \hat{\theta}_k^{(1:T)}
		\end{pmatrix} \alpha , 
		\]
		where $\otimes$ denotes the Kronecker product.
		Let $\boldx_i = \begin{pmatrix} x_{i}^{(1)} \otimes \hat{\theta}_1^{(1:T)} & \cdots & x_{i}^{(k)} \otimes \hat{\theta}_k^{(1:T)}
		\end{pmatrix} \trans $ be the $k \times T$ design matrix for observation $i$ and let $\boldx_{\mcC}$ be the set of all such matrices for $i \in \mcC$. We will write $\boldx_i^{(t)}$ to denote the $t^\text{th}$ column of $\boldx_i$, which is equal to $\begin{pmatrix} x_{i}^{(1)} \hat{\theta}_1^{(t)} & \cdots & x_{i}^{(k)} \hat{\theta}_k^{(t)} \end{pmatrix}\trans$.
		
		Rewriting the part of equation (\ref{eq:slimp_loss}) to the right of the infimum operator in terms of this new variable, we get
		\[
		\sum_{t=1,t' = 1}^{T} \|\hat{\mu}_i^{(t)} - (\boldx_i^{(t')}) \trans \alpha \|_2^2 \gamma_{t,t'}.
		\]
		We note that this looks like a weighted linear regression. Expanding the quadratic term we obtain
		\begin{align*}
			\sum_{t=1,t' = 1}^{T} \|\hat{\mu}_i^{(t)} - (\boldx_i^{(t')}) \trans \alpha \|_2^2 \gamma_{t,t'}&= \sum_{t=1,t' = 1}^{T}  \left[(\hat{\mu}_i^{(t)})^2 \gamma_{t,t'}  - 2 \alpha \trans \boldx_i^{(t')} \hat{\mu}_i^{(t)}  \gamma_{t,t'} + \alpha\trans \boldx_i^{(t')} (\boldx_i^{(t')} )\trans \alpha \gamma_{t,t'}  \right]\\
			&= \sum_{t=1}^{T}  \left[(\hat{\mu}_i^{(t)})^2 \sum_{t' = 1}^{T} \gamma_{t,t'}  \right]+ \sum_{t' = 1}^{T}  \left [\alpha\trans \boldx_i^{(t')} (\boldx_i^{(t')} )\trans \alpha \sum_{t=1}^{T}\gamma_{t,t'}\right]  \\
			& \qquad \qquad - 2 \sum_{t=1,t' = 1}^{T} \alpha \trans \boldx_i^{(t')} \hat{\mu}_i^{(t)}  \gamma_{t,t'}\\
			&=\frac{\sum_{t=1}^{T}  (\hat{\mu}_i^{(t)})^2 }{T} + \frac{\sum_{t' = 1}^{T} \alpha\trans \boldx_i^{(t')} (\boldx_i^{(t')} )\trans \alpha }{T} - 2 \sum_{t=1,t' = 1}^{T} \alpha \trans \boldx_i^{(t')} \hat{\mu}_i^{(t)}  \gamma_{t,t'}\\
			&=\frac{\hat{\mu}_i\trans \hat{\mu}_i} {T} + \frac{1}{T}\alpha\trans \boldx_i \boldx_i\trans \alpha  - 2 \sum_{t=1,t' = 1}^{T} \alpha \trans \boldx_i^{(t')} \hat{\mu}_i^{(t)}  \gamma_{t,t'}\\
		\end{align*}
		Focusing on the part depending on $\alpha$, we see that we have the quadratic form
		\[ 
		\alpha \trans \Sigma_{\boldx, \boldx}  \alpha - 2 \alpha \trans \Sigma_{\boldx, \hat{\mu}},
		\]
		where
		\begin{align*}
			\Sigma_{\boldx, \boldx} &= \frac{1}{N \cdot T}   \sum_{i =1}^N \boldx_i  \boldx_i \trans
			\intertext{and} 
			\Sigma_{\boldx, \hat{\mu}} &= \frac{1}{N}   \sum_{i =1}^N   \sum_{t =1,t' =1}^T \boldx_i^{(t')}\hat{\mu}_i^{(t)} \gamma_{t,t'}.
		\end{align*}
		The additional scaling by $1/N$ is mathematically unnecessary but helps provide some computational stability.
	\end{pf}
	
	\clearpage
	
	\section{Additional plots} \label{sec:extragraph}
	
	\subsection{Toy example}
	This section presents additional plots from Section \ref{sec:toy} for $\rho = 0.0$ and $\rho = 0.9$.
	
	
	\begin{figure}[H]
		\includegraphics[width=\textwidth]{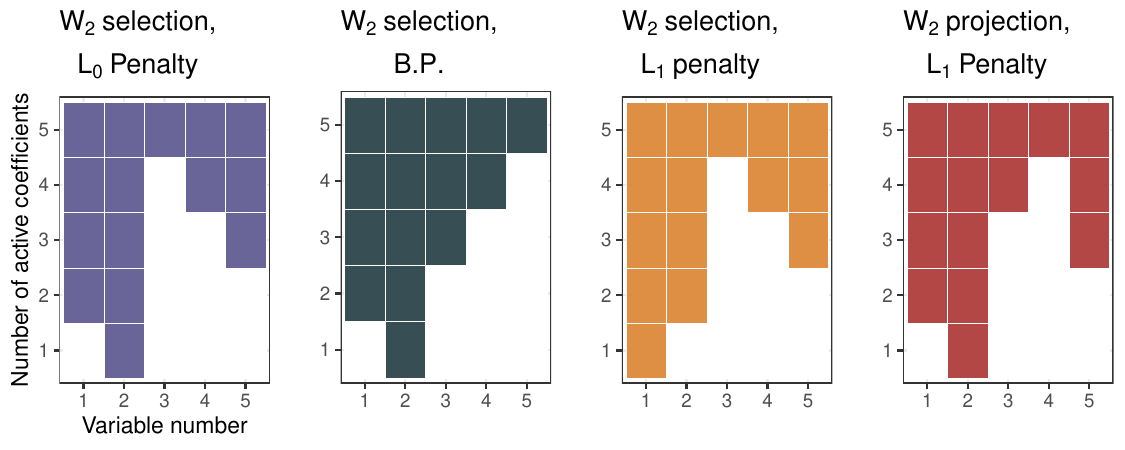}
		\caption
		{\textbf{Selection order for SLIM-p model with data correlation $\mathbf{\rho = 0.0}$.} We ran best subsets (left panel) and relaxed quadratic binary program using an $L_1$ like penalty ( (MCP net penalty with elastic net parameter 0.99 and MCP parameter 1.5)) like in equations (\ref{eq:alpha_unconst}) and (\ref{eq:selection}) using a penalized regression (middle panel), and for an exact binary quadratic program (right panel). }
	\end{figure}
	
	\begin{figure}[h]
		\includegraphics[width=\textwidth]{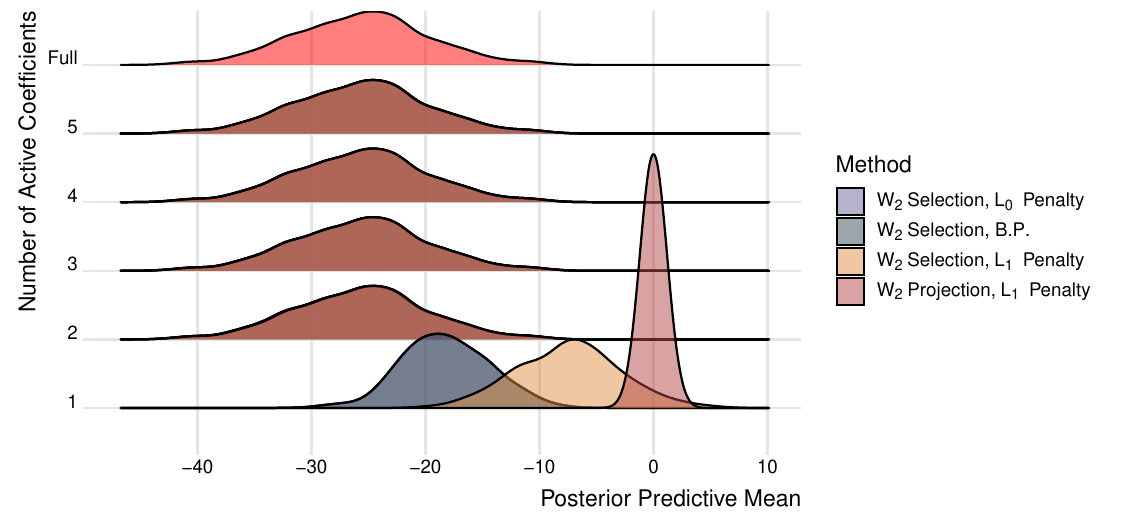}
		\caption
		{\textbf{Ridge plots for $\mathbf{\rho = 0.0}$.} The left panel is a ridge plot for the SLIM-a estimation and the right panel is a ridge plot for the SLIM-p estimation for the toy example in Section \ref{sec:toy}. }
	\end{figure}
	
	
	\begin{figure}[H]
		\includegraphics[width=\textwidth]{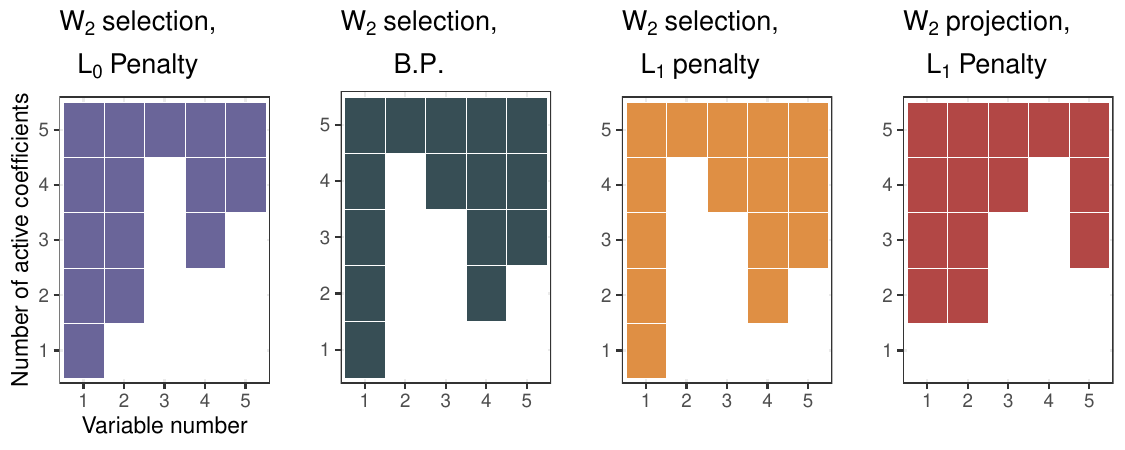}
		\caption
		{\textbf{Selection order for SLIM-p model with data correlation $\mathbf{\rho = 0.9}$.} We ran best subsets (left panel) and relaxed quadratic binary program using an $L_1$ like penalty ( (MCP net penalty with elastic net parameter 0.99 and MCP parameter 1.5)) like in equations (\ref{eq:alpha_unconst}) and (\ref{eq:selection}) using a penalized regression (middle panel), and for an exact binary quadratic program (right panel). }
	\end{figure}
	
	\begin{figure}[!h]
		\includegraphics[width=\textwidth]{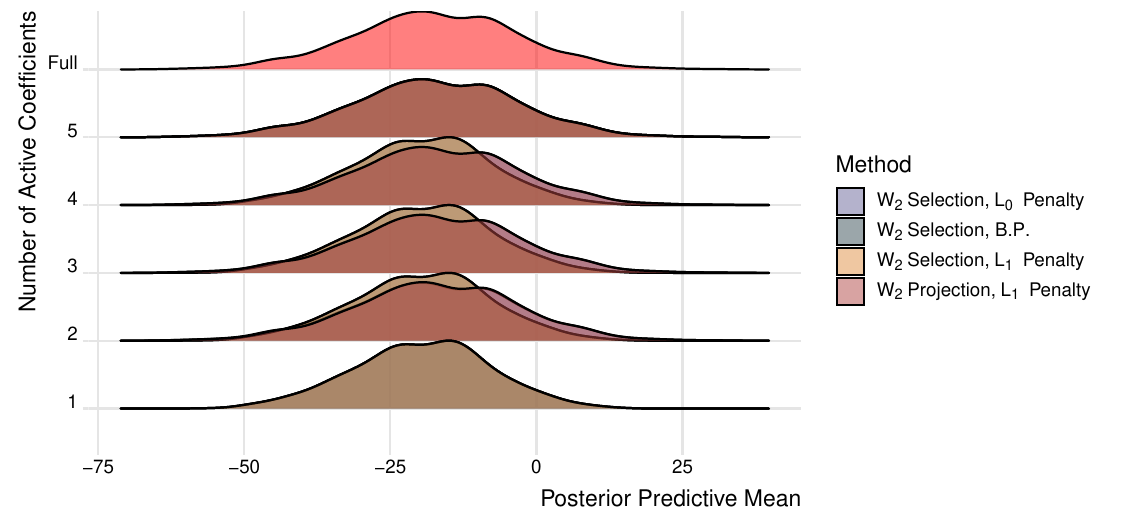}
		\caption
		{\textbf{Ridge plots for $\mathbf{\rho = 0.9}$.} The left panel is a ridge plot for the SLIM-a estimation and the right panel is a ridge plot for the SLIM-p estimation for the toy example in Section \ref{sec:toy}. }
	\end{figure}
	
	\clearpage

	\clearpage
	
	\section{Additional plots for applied data analysis}\label{sec:data_graph}
	\textbf{Curated Ovarian Data}\\
	\begin{figure}[!htb]
		\centering
		\includegraphics[width = \textwidth]{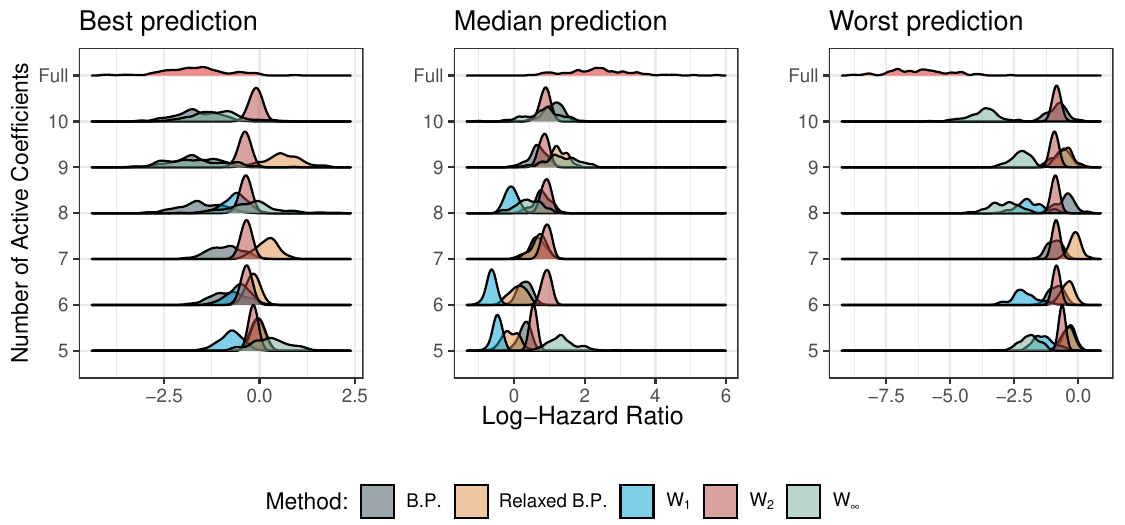}
		\caption{\textbf{Ridgeplots on the validation data.} The plots detail the best, median, and worst average 2-wasserstein predictions on the validation data from the curateOvarianData set. The quality of prediction is chosen by examining the distance from the prediction from the full model ($\eta$) for a given model size. The evaluation model size was chosen as 10 covariates but we also include down to 5 covaraites to see how predictions change as one adds more covariates. The top line in the ridge plots are the predictions from the full model including all covariates.}
		\label{fig:ovar_train_ridge}
	\end{figure}
	\clearpage
	\textbf{Glioblastoma Data}\\
	\begin{figure}[!bth]
		\centering
		\includegraphics{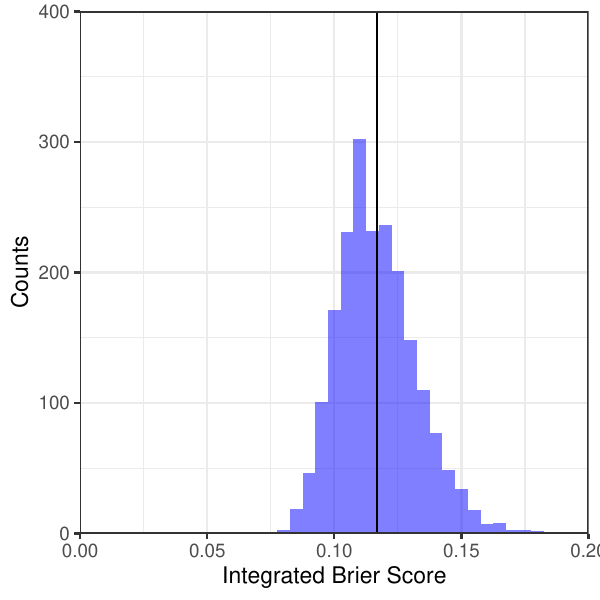}
		\caption{Distribution of the integrated Brier score for the AFT survival model described in Section \ref{sec:gbm_sec} over the draws from the posterior. The integrated Brier score is evaluated on the validation set. The line denotes the expectation of the samples}
		\label{fig:int_bs}
	\end{figure}
	\begin{figure}[!bth]
		\centering
		
		\includegraphics{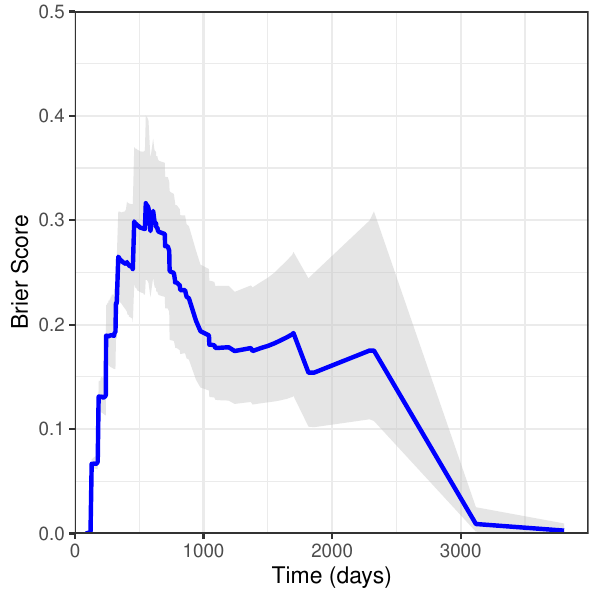}
		\caption{Brier score over time for the AFT model in Section \ref{sec:gbm_sec}. Line denotes the mean Brier score and the bands denote the 95\% credible intervals. Brier score was evaluated on the validation set. We can see that the model does a poor job for some early times but does better towards the beginning and end.}
		\label{fig:bs_time}
	\end{figure}

	\vskip 0.2in
	
\end{document}